\documentclass[10pt,aps,prb,showpacs,superscriptaddress,twocolumn,floatfix]{revtex4}
\usepackage{amsmath}
\usepackage{citesort}
\usepackage{graphicx}
\usepackage{dcolumn}
\begin{document}
\title{Optical Sum Rule in Finite Bands}
\author{J.P. Carbotte}
\affiliation{Department of Physics and Astronomy, McMaster University,\\
Hamilton, Ontario, L8S 4M1 Canada}
\author{E. Schachinger}
\email{schachinger@itp.tu-graz.ac.at}
\homepage{www.itp.tu-graz.ac.at/~ewald}
\affiliation{Institute of Theoretical and Computational Physics\\
Graz University of Technology, A-8010 Graz, Austria}
\date{\today}
\begin{abstract}
In a single finite electronic band the total optical spectral weight
or optical sum carries information on the interactions involved
between the charge carriers as well as on their band structure.
It varies with temperature as well as with impurity scattering.
The single band optical sum also bears some relationship to the charge
carrier kinetic energy and, thus, can potentially provide
useful information, particularly on its change as the charge
carriers go from normal to superconducting state. Here we
review the considerable advances that have recently been made in
the context of high $T_c$ oxides, both theoretical and experimental.
\end{abstract}

\maketitle
%\newpage

\section{Introduction}

Much is known about the properties of the superconducting state
in the cuprates, yet after 20 years of intense research there is,
as yet, no consensus as to the driving interactions responsible
for the pairing (mechanism). It is well established that the
superconducting order parameter has $d$-wave rather than
conventional $s$-wave symmetry. Angular resolved photoemission
spectroscopy\cite{ref1,ref2} (ARPES) finds that the leading
edge of the electron spectral density at the Fermi energy shifts
below the chemical potential as a result of the onset of superconductivity.
The observed shift is zero in the nodal direction $(\pi,\pi)$
and maximum in the antinodal direction $(0,\pi)$.
More recently a non monotonic increase of
the gap amplitude as a function of direction has been observed
in an electron doped cuprate but the symmetry is again $d$-wave%
.\cite{ref3}

Many other independent experimental techniques have provided
additional compelling evidence for $d$-wave symmetry. One such
technique is microwave spectroscopy which gives a linear in
temperature $(T)$ dependence of the superfluid density at low
$T$ [\onlinecite{ref4}]. Complimentary to the above techniques are
phase sensitive experiments\cite{ref5} based on Josephson
tunneling and flux quantization which provide direct evidence for a
change in sign of the order parameter not just that it has a zero.

Another important observation which points directly to a non
conventional mechanism is the so called collapse of the
inelastic scattering rate\cite{ref6,ref7,ref8,ref9} in the
superconducting state which results in a large peak at intermediate
$T$ below the critical temperature $(T_c)$ in the real part of the
microwave conductivity.\cite{ref6,ref7,ref10,ref11} The origin of this
peak is understood in terms of a readjustment in the excitation spectrum
involved in the interaction as superconductivity develops. This
is expected in electronic mechanisms\cite{ref6,ref9} involving
the spin or charge susceptibility which looses weight at small
energies corresponding to a hardening of the spectrum and little
inelastic scattering remains at low $T$.

All the evidence reviewed above points to deviations from a simple
BCS $s$-wave phonon driven superconductivity, but, nevertheless,
the search goes on for other essential differences which could
help identify the underlying mechanism. An avenue to explore is
the idea of kinetic as opposed to potential energy driven
superconductivity. The idea is that it is a reduction in kinetic
energy (KE) that drives the condensation into the superconducting state
in contrast to BCS theory where the potential is decreased
sufficiently to overcome a KE increase and provides as well a
reduction in total energy. The possibility of KE driven
superconductivity was considered by Hirsch\cite{ref12} and
explicitely demonstrated for the hole mechanism of
superconductivity\cite{ref13,ref14,ref15} for which the effective
mass of the holes decreases in the superconducting state by
pairing. It is also expected in other theoretical frameworks
for highly correlated systems.\cite{ref16,ref17}

In a single finite tight binding band with nearest neighbors
hopping only, the KE
is related to the optical sum defined as the total optical spectral
weight under the real part of the optical conductivity
$\sigma_1(T,\omega)$ integrated over energy $\omega$.\cite{ref19} This fact
applies whatever the interactions involved, the temperature,
and also when impurities are present. Of course, for this to hold
the conductivity needs to be computed exactly including vertex
corrections. At first sight the
restriction to a tight binding band with nearest neighbors only
appears restrictive but in reality it has been found that KE
and optical sum (OS) track each other reasonably well
even when higher neighbors are considered (second, third, etc.).
In view of this fact several recent experimental and theoretical
papers have appeared concerning the temperature dependence of the
OS in the normal state and its change in the superconducting state.
Here, we review this body of work with the aim of understanding
what it tells us about the underlying interactions involved.
We will restrict our discussion to the cuprates and to their
in-plane response. For the $c$-axis motion a sum rule
violation was noticed by Basov {\it et al.}\cite{ref18}
but the KE involved in the $c$-axis motion is small.\cite{ref20,ref21}
To investigate its relation to the total condensation energy
of the transition to the superconducting state, it is necessary to
consider the $ab$-plane of the cuprates. Exactly how to treat the
$c$-axis transport also complicates the interpretation. Some
possible mechanisms include strong intra layer scattering,\cite{ref22}
non Fermi liquid ground states,\cite{ref23} confinement,\cite{ref16}
inter-plane and in-plane charge fluctuations,\cite{ref24} indirect
$c$-axis coupling through the particle-particle channel,\cite{ref25}
resonant tunneling on localized states in the blocking layer,\cite{ref26}
two band models,\cite{ref27,ref28} and coherent and incoherent
tunneling.\cite{ref29,ref30,ref31,ref32,ref33} Clearly this problem
is worth further study but is not part of this review which will deal
only with the cuprates and their $ab$-plane response.

The in-plane optical sum (OS) in the cuprates is observed to increase
as the temperature is decreased below room temperature.
In the normal phase it is often, but not always, found to
follow a $T^2$ law. The changes are
small but larger than is expected on a rigid band model neglecting
interactions. % and often, but not always, found to follow a $T^2$ law.
In the superconducting phase, on the other hand,
%Below the critical temperatureRe: Schachinger - Weber
a change in slope of this $T^2$
behavior is seen. While there remain some differences in details
between the various experimental groups, for underdoped samples it is
agreed that the OS falls above the extrapolated normal state value.
This is clearly seen in Fig.~\ref{fig:14} which was reproduced from
van der Marel {\it et al.}\cite{ref34} and deals with
Bi$_2$Sr$_2$CaCu$_2$O$_{8+x}$
(BSCCO, Bi2212). This behavior is opposite to that expected on the
basis of BCS theory for which the OS is predicted to decrease. This
fact can be traced to an increase in KE due to the
formation of Cooper pairs. On the other hand, such a conventional
behavior of the OS or change in KE on entering the superconducting
state was observed in overdoped samples of BSCCO by Deutscher
{\it et al.}\cite{ref87a} Their data is reproduced in our
Fig.~\ref{fig:19}. These authors also find that the crossover
from negative to positive KE change occurs around, but slightly
above optimum doping (see Fig.~\ref{fig:20}). These findings
have been largely confirmed by Carbone {\it et al.}\cite{ref33a} It is
these facts that this review is directly concerned with and seeks
to understand. Theoretically we will find that the temperature
dependence of the OS can in some cases be dominated by a term which is
proportional to a specific average over energy of the quasiparticle
inelastic scattering rates and this need not give a $T^2$ law. This
term is missing in all theories that do not explicitly treat damping
effects. Further, this non $T^2$ dependence has implications for the
accuracy of experimental results on the OS difference between
superconducting and normal state which require an extrapolation to
low temperatures of normal state data known only above $T_c$.

The paper is structured as follows. We begin with theoretical
considerations and then review experimental information as it relates
to the calculations.
Section \ref{ssec:2a} introduces the optical sum and its relation to the
kinetic energy. In
Sec.~\ref{ssec:2d} we provide simple analytic formulas for the KE and
OS in a free electron model which allows one to understand some
qualitative aspects of this relationship. It also contains a general
formulation of the OS for a
simplified tight binding model which averages over anisotropies and
greatly simplifies the mathematics. We argue that the temperature
dependence of the OS is governed mainly by the inelastic scattering.
Both, real and imaginary part of the charge carrier self energy are
important. In Sec.~\ref{ssec:2b} we describe the Nearly
Antiferromagnetic Fermi Liquid model (NAFFL), give the set of generalized
Eliashberg equations needed for numerical work, and also specify the
electron-spin fluctuations interaction (MMP model). We give results for the
temperature dependence of the OS and of the KE in the normal and
superconducting state. In Sec.~\ref{ssec:2c} we introduce the Hubbard
model and Dynamical Mean Field Theory (DMFT) and give results. 

In Sec.~\ref{ssec:3b} we summarize some of the important
effects a finite band cutoff has on the self energy. Section~\ref{ssec:3c}
is devoted to results for the OS when a delta function, and
Sec.~\ref{ssec:3d} when an MMP form
(spin fluctuations) is used for the electron-boson interaction.
Section~\ref{ssec:3e} provides an analysis of the temperature
dependence of the OS for coupling to a low energy boson which
results in a linear
rather than quadratic in $T$ law. It is argued that other temperature
dependences could arise when different model interactions are used.

In Sec.~\ref{ssec:4a} we investigate within the NAFFL model the effect on
the OS of the collapse of the inelastic scattering on entering the
superconducting state which can lead to an increase in the OS rather than
the usual BCS behavior. We also present experimental results in the
BSCCO cuprates. Section~\ref{ssec:4c} deals
with a simplified qualitative model based on a temperature dependent
scattering rate
which decreases with $T$ as $T^4$. While this model is not
accurate, it shows clearly how the KE in the superconducting state can
decrease below its normal state value due to the collapse of the
inelastic scattering.
In Sec.~\ref{ssec:4b} we describe a related,
more phenomenological model due to Norman and P\'epin which has several
common elements with the model discussed in
Sec.~\ref{ssec:4a}. We also provide comparison with
experiment and additional theoretical results for the KE change on
condensation into the superconducting state in Sec.~\ref{ssec:4e}.
Section~\ref{ssec:4d}
gives DMFT results for the normal as well as superconducting state
in the $t$-$J$ model. KE and potential energy are discussed and
compared with experiment. Further results based on the Hubbard
model are commented on as are those based on the negative $U$
Hubbard model
used to describe the BCS - BE (Bose - Einstein) crossover.

Section~\ref{sec:5} deals with models of the pseudogap state above
$T_c$ (the superconducting critical temperature) that exists up
to a temperature $T^\ast>T_c$, with $T^\ast$ the pseudogap temperature.
In Sec.~\ref{ssec:5a} we discuss KE changes in the preformed pair model
in which pairs form at the pseudogap temperature $T^\ast$ and
superconductivity sets in only at a lower temperature when phase
coherence is established. In Sec.~\ref{ssec:5b} we consider an alternative
model to phase fluctuations of the pseudogap state, namely the
$D$-density wave model, a competing ordered state with bond currents
and associated magnetic moments. In Sec.~\ref{sec:6} we deal briefly with
the problem of spectral weight distribution as a function of energy.
A short summary is provided in Sec.~\ref{sec:7}.

\section{THEORY}
\label{sec:2}
\subsection{General considerations}
\label{ssec:2a}

The single band OS is defined as\cite{ref34}
\begin{equation}
  \label{eq:1}
  \int\limits_{-\Omega}^\Omega\!d\omega\,\Re{\rm e}[
  \sigma_{1xx}(\omega)] = \frac{\pi e^2}{\hbar^2V}
  \sum\limits_{{\bf k},\sigma}n_{{\bf k},\sigma}
  \frac{\partial^2\varepsilon_{\bf k}}{\partial k_x^2} =
  \frac{\pi e^2}{\hbar^2} W.
\end{equation}
Here $\sigma_{1xx}(\omega)$ is the real part of the 
$(x,x)$-component of the optical conductivity,
$e$ is the electron charge, $\hbar$ Planck's constant, {\bf k}
momentum, $\sigma$ spin, and $V$ the crystal volume. The limits
$-\Omega$ to $\Omega$ on the conductivity integral are to be taken
to include all the contributions to $\sigma_{1xx}(\omega)$ from the
band of interest and from no other.
The integration from $-\Omega$ to $\Omega$
can be restricted to $(0,\Omega)$ by making use of the fact that
the real part of the conductivity is an even function of $\omega$.
Often the integration on $\Omega$ is also extended to infinity under
the condition that only transitions from the one band of interest
are included.
The sum over {\bf k} extends over the
entire first Brillouin zone, $n_{{\bf k},\sigma}$ is the probability
of occupation of a state $\vert{\bf k},\sigma\rangle$, and
$\varepsilon_{\bf k}$ is the electron dispersion relation.
$W$ is the OS divided by $\pi e^2/\hbar^2$ and will be quoted in
meV. In deriving this OS rule the one body part of the Hamiltonian
gives the single particle orbitals of energy
$\varepsilon_{\bf k}$ and the two body piece
provides correlation effects beyond those included in
\begin{table}[tp]
\caption{\label{tab:1}The two tight binding models used within this
paper. Model A corresponds to the tight binding model discussed by
van der Marel {\it et al.}\protect{\cite{ref34}} $t$ and $t'$ are
given in meV, the critical temperature $T_c$ in K, and the filling
$\langle n\rangle$ is defined in Eq.~\protect{\eqref{eq:8}}.
}
\begin{ruledtabular}
\begin{tabular}{ldddd}
Model & t & t' & \langle n\rangle & T_c\\
\hline
A & 148.8 & 40.9 & 0.425 & 90\\
B & 100.0 & 16.0 & 0.4 & 100\\
\end{tabular}
\end{ruledtabular}
%\vspace*{10cm}
\end{table}
$\varepsilon_{\bf k}$ (referred to as kinetic energy even though it
can contain effects of interactions).
If all bands are included rather than just the one in Eq.~\eqref{eq:1}
one would obtain the exact sum rule proportional to $n/m$ with $m$
the bare electron mass and $n$ the total electron density. This
is a constant independent of temperature. It does not change when
the system undergoes a phase transition to the superconducting state
by charge conservation. This also applies to the free electron
infinite band and is the basis for the Ferrell-Glover-Tinkham
sum rule\cite{ref123,ref124} which will be elaborated upon in Sec.~\ref{sec:6}.

\begin{figure}[tp]
%\centering
  \vspace*{-4mm}
  \includegraphics[width=9cm]{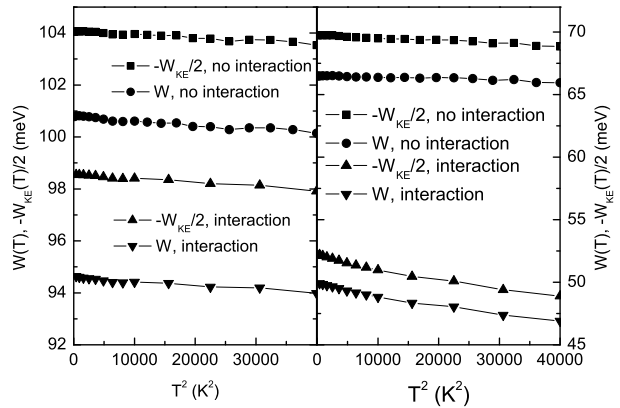}
\vspace*{-8mm}
  \caption{Optical sum $W$ and kinetic energy, $-W_{\rm KE}/2$,
as a function of $T^2$. Solid circles and squares are for the non
interacting case while solid up-triangles and solid down-triangles
include interactions.
The left hand frame applies to
Model A of Table~\protect{\ref{tab:1}} and an MMP model
parameter $\omega_{SF}=82\,$meV [see Eq.~\protect{\eqref{eq:5}}] was
used. Here, the interacting and non interacting cases show similar
temperature dependencies. The right hand frame is for
Model B of Table~\protect{\ref{tab:1}} with an MMP model parameter
$\omega_{SF}=10\,$meV. Note the difference in temperature dependence
between interacting and non interacting case.
}
  \label{fig:1}
\end{figure}
The
important observation for this review is that for tight binding bands
with nearest neighbors hopping, $t$, only, i.e.:
\begin{equation}
  \label{eq:2}
  \varepsilon_{\bf k} = -2t\left[\cos(ak_x)+\cos(ak_y)\right],
\end{equation}
where $a$ is the CuO$_2$ plane lattice parameter,
\begin{equation}
  \label{eq:3}
  W\equiv-\frac{1}{2}W_{KE}\equiv -\frac{1}{2}
  \frac{a^2}{V}\sum\limits_{{\bf k},\sigma} n_{{\bf k},\sigma}
  \varepsilon_{\bf k}.
\end{equation}
$W_{KE}$ is the kinetic energy per copper atom.
Thus, for this simple case an experimental measurement of $W$
gives a measure of the KE. The band structure in the oxides is not,
in general, limited to first neighbor hopping and one may well
wonder if the above observation is of any practical use.

For second neighbor hopping ($t,\, t'$ model)
\begin{eqnarray}
  \varepsilon_{\bf k} &=& -2\left\{
   t\left[\cos(ak_x)+\cos(ak_y)\right]\right.\nonumber\\
  &&\left.- 2t'\cos(ak_x)\cos(ak_y)
   \right\} - \mu, 
  \label{eq:4}
\end{eqnarray}
with $\mu$ the chemical potential. In Fig.~\ref{fig:1} we
compare results for $W$ (solid circles) with results for
$-W_{KE}/2$ (solid squares) for two models, A and B, left and right hand
frame, respectively. Parameters of the models are defined in
Table~\ref{tab:1} where $\langle n\rangle$ is the filling  factor
which determines the chemical potential; with $\langle n\rangle = 0.5$,
half filling, $\mu = 0$ and $t'=0$ in Eq.~\eqref{eq:4}. No interactions are
included beyond those giving rise to the tight binding band structure.
It is clear that while $W$ in both cases is a few percent
smaller than $-W_{KE}/2$ their dependence on temperature track
each other fairly well. Thus, it makes sense to pursue this avenue as
a means to get information on the KE and its variation with temperature
and also when it undergoes a phase transition to the superconducting
state. One should be aware, however, that for some sets of tight
binding parameters the difference between OS and KE can be much larger
and they may no longer track each other (see the very recent
paper by Marsiglio {\it et al.}\cite{ref124a} for more discussion).
A final point should be stressed. To compute the OS in Eq.~\eqref{eq:1}
it is not necessary, although one can if one wishes, to calculate
the conductivity which depends on a two-particle Green's function
and is to be calculated with vertex corrections. The right hand side
of Eq.~\eqref{eq:1}, instead, depends only on the one particle
spectral density $n_{{\bf k},\sigma}$ which determines the probability
of occupation of the state $\vert\textbf{k},\sigma\rangle$.
On the other hand, if one is interested in the optical spectral
weight integrated only to $\omega_M$ with $\omega_M<\Omega$ one
can no longer avoid calculating the conductivity. For this reason
the problem of optical weight distribution is left to a last
section \ref{sec:6}.

\subsection{The Continuum Limit and Quadratic Dispersion; Simple Results}
\label{ssec:2d}

To get some insight into the significance of the OS, we
consider next several simplifications. The probability of occupation
of the state $\vert{\bf k}\rangle$ (both spins)
is related to the charge carrier
spectral density $A({\bf k},\omega)$ by
\begin{equation}
  \label{eq:10}
  n_{\bf k}(T) = \int\limits_{-\infty}^\infty\!d\omega\,f(T,\omega)
  A({\bf k},\omega),
\end{equation}
where $f(T,\omega)$ is the Fermi-Dirac distribution function at
temperature $T$.
For tight binding electrons with no correlation effects the spectral
function $A({\bf k},\omega)=\delta(\varepsilon_{\bf k}-\omega)$ and
$n_{\bf k}(T) = f(\varepsilon_{\bf k},T)$. Before proceeding to
include interactions it is convenient in what follows to rewrite
Eq.~\eqref{eq:1} in an equivalent form through integration by parts
with the surface integral equal to zero by symmetry in a periodic
band:
\begin{equation}
  \label{eq:11}
  \bar{W} = -\frac{2}{V}
  \sum\limits_{\bf k}\left(\frac{\partial\varepsilon_{\bf k}}{\partial
  {\bf k}}\right)^2\frac{dn_{\bf k}}{d\varepsilon_{\bf k}}.
\end{equation}

Knigavko {\it et al.}\cite{ref49} have found this second form to be
very useful when making approximations. In the Kubo formula which
defines the conductivity $\sigma_{1xx}(\omega)$ which appears on the
left hand side of Eq.~\eqref{eq:1} it is the velocity squared,
$(\partial\varepsilon_{\bf k}/\partial k_x)^2$ that enters naturally.
In making approximations to the band structure in order to get
simplified expressions it is important to have the same quantity
$(\partial\varepsilon_{\bf k}/\partial k_x)^2$ on both sides of
Eq.~\eqref{eq:1} and, indeed, in Ref.~\onlinecite{ref49} it is verified
that the right hand side (RHS)
of Eq.~\eqref{eq:1} and $\bar{W}$ of Eq.~\eqref{eq:11} agree
to high accuracy for the two simplified band structure models they
considered, namely free electron bands cut off to $-D/2$ and $D/2$ at
half filling and another model designed to treat the tight binding case
which we will define later. In both cases the density of states was
approximated by a constant. Such an approximation, however, is not
completely compatible with the assumed Fermi velocity model and this
explains why we use $\bar{W}$ rather than $W$.
In the continuum limit of Eq.~\eqref{eq:4}
with $t' = \mu = 0$ as the lattice parameter $a\to 0$ one gets
\[
  \varepsilon_{\bf k} = -4t+\frac{\hbar^2}{2m}\left(k_x^2+k_y^2\right),
\]
with $\hbar^2/2m = t a^2$ (where this product is assumed to be finite
as $a\to 0$) or $a^2 = \frac{\hbar^2}{m}\frac{4}{D}$
and $D/2 = 4t$. For these approximations
\[
  W = \bar{W} = \frac{(\hbar\Omega_p)^2}{4\pi e^2},
\]
the well known result for free electron bands, with $\Omega^2_p =
4\pi ne^2/m$ the plasma frequency squared and $n$ the electron density.
We can also show that
\[
  W_{KE} = -W,
\]
so that the relationship between KE and OS found for
tight binding bands with first neighbors is profoundly changed
when these are approximated by their continuum
limit (free electron case). This is, in a sense, the extreme opposite
case to the nearest neighbor only tight binding model.
 These results were obtained at zero
temperature. At finite $T$ things are not quite as simple. $W$
and $\bar{W}$ undergo no change with temperature but
$W_{KE}$ does and is
\[
  W_{KE} = \frac{N(0)D\hbar^2}{m}\left[\frac{\pi^2}{3}\left(
  \frac{k_BT}{D/2}\right)^2-1\right],
\]
which is the well known result that the KE increases as the square of
the temperature normalized to the half band width $D/2$. These results
show that the correspondence between $W$ and $W_{KE}$ can be
rather subtle and it can be lost when approximations to the tight
binding band are made. Here $N(0)$ is the charge carrier density
of states at the Fermi energy.

Knigavko {\it et al.}\cite{ref49} used a
somewhat more sophisticated model to approximate a tight binding
band. This model was used previously by Marsiglio and Hirsch.\cite{ref50}
In this approach the square of the electron velocity is replaced by
\[
%  \label{eq:14}
  \left(\frac{1}{\hbar}\frac{\partial\varepsilon_{\bf k}}
  {\partial k_x}\right)^2 =
  \frac{D}{4m_b}\left[1-\left(\frac{\varepsilon_{\bf k}}{D/2}
  \right)^2\right]
\]
and
\begin{equation}
  \label{eq:15}
  \bar{W}(T) = \frac{\hbar^2}{m_b}\left(-\int\limits_{-\infty}^\infty
  \frac{d\omega}{D/2}\,f(T,\omega)
  \int\limits_{-D/2}^{D/2}\frac{d\epsilon}{D/2}\,
  \epsilon A(\epsilon,\omega)\right).
\end{equation}
Here, $m_b$ is a band electronic mass. In this model the rule that
$\bar{W} = -W_{KE}/2$ still holds. 
For $A(\epsilon,\omega) = \delta(\epsilon-\omega)$ (no correlations)
we recover
\begin{equation}
  \label{eq:16}
  \bar{W}(T) = \frac{\hbar^2}{m_b}\left[1-\frac{1}{3}\left(
  \frac{k_BT}{D/2}\right)^2\right],
\end{equation}
which is a result obtained by van der Marel {\it et al.}\cite{ref34}
and often used to interpret experiments. It is important to contrast this
result with the free electron case for which
\[
 \left(\frac{1}{\hbar}\frac{\partial\varepsilon_{\bf k}}{\partial
 k_x}\right)^2 = \frac{D/2+\varepsilon_{\bf k}}{m}
\]
and
\begin{equation}
  \label{eq:16a}
  \bar{W}(T) = \frac{\hbar^2 n}{m}\left[1-\frac{2}{n}
  \int\limits_{-\infty}^\infty\!d\omega\,f(T,\omega)A(D/2,\omega)
  \right],
\end{equation}
where $n$ is the charge carrier density.

To treat interactions it is convenient to rewrite Eq.~\eqref{eq:15}
in the form
\begin{equation}
  \label{eq:17}
  \bar{W}(T) = \frac{\hbar^2}{2m_b}\left[1-\int\limits_{-\infty}^\infty
  \!d\omega\,f(T,\omega)h(T,\omega)\right],
\end{equation}
where
\begin{equation}
  \label{eq:18}
  h(T,\omega) = \frac{4}{(D/2)^2}\int\limits_0^{D/2}\!d\epsilon\,
  \epsilon A(\epsilon,\omega).
\end{equation}
Note also that the probability of occupation of the state $\epsilon$
given in Eq.~\eqref{eq:10}
%\begin{equation}
%  \label{eq:18a}
%  n(\epsilon,T) = \int\limits_{-\infty}^\infty\!d\epsilon\,
%  f(T,\epsilon)A(\epsilon,\omega)
%\end{equation}
is closely related to Eqs.~\eqref{eq:17} and \eqref{eq:18}. We can
apply the Sommerfeld expansion to Eq.~\eqref{eq:18} to get
\begin{eqnarray}
  \bar{W}(T) &=& \frac{\hbar^2}{2m_b}\left[1-\int\limits_{-\infty}^0\!
  d\omega\,h(T,\omega)-\frac{\pi^2}{6}\left(k_BT\right)^2\right.
  \nonumber \\
  && \times\left.
  \left.\frac{d\,h(T,\omega)}{d\,\omega}\right\vert_{\omega=0}
  \right].
  \label{eq:19}
\end{eqnarray}
For no interactions $h(T,\omega) = 4\omega/(D/2)^2$ for $\omega\ge 0$
and zero for $\omega<0$
[note that the derivative in Eq.~\eqref{eq:19} (non interacting case)
is to be weighted by $1/2$ at $\omega=0$]
so that the second term in Eq.~\eqref{eq:19}
vanishes and the third term
gives the first correction in \eqref{eq:16}. In terms of the real and
imaginary part of the self energy we can work out $h(T,\omega)$ to
get
\begin{eqnarray}
  h(T,\omega) &=& \frac{4}{\left(\frac{D}{2}\right)^2}\left\{\frac{\chi}{\pi}
  \left[\tan^{-1}\left(\frac{\frac{D}{2}-\chi}{y}\right)+
  \tan^{-1}\left(\frac{\chi}{y}\right)\right]\right.\nonumber\\
  &&\left.+
  \frac{y}{2\pi}\ln\left\vert\frac{y^2+\left(\frac{D}{2}-\chi\right)^2}
  {y^2+\chi^2}\right\vert\right\},
  \label{eq:20}
\end{eqnarray}
where $\chi=\omega-\Sigma_1(\omega)$ and $y = -\Sigma_2(\omega)$ with
$\Sigma_1(\omega)$ [$\Sigma_2(\omega)$] the real [imaginary] part of
the self energy $\Sigma(\omega)$. If the real part of $\Sigma$ is
neglected and its imaginary part is assumed to be a constant $\Gamma$
as it would be for impurities in an infinite band we can work out the
integral in Eq.~\eqref{eq:19} and get:
\begin{equation}
  \label{eq:21}
  \bar{W}(T) = \frac{\hbar^2}{2m_b}\left[1-\frac{8\Gamma}{D\pi}-
  \frac{\pi^2}{3}\left(\frac{k_BT}{D/2}\right)^2\right].
\end{equation}
Note that the second term which deals directly
with interactions between
the electrons is of order some scattering rate over $D/2$ while the
third which has often been emphasized in literature, is of order
$[k_BT/(D/2)]^2$ and, therefore, can be expected to be much smaller
than the first. In the oxides, as an example, $\Gamma$ is known to be
of order $T$.\cite{ref126}
Eq.~\eqref{eq:21} was arrived at independently by
Benfatto {\it et al.}\cite{ref52} and by Karakozov and Maksimov.\cite{ref53}
It is central to any discussion of single band sum rule. While the
approach taken in Refs.~\onlinecite{ref52} and \onlinecite{ref53} are
somewhat different much of the basic content is similar.

Finally, we consider the superconducting state at $T=0$ for which in
BCS theory
\begin{equation}
   \label{eq:11a}
   n_{\bf k} = \frac{1}{2}\left(1-\frac{\varepsilon_{\bf k}}
   {\sqrt{\varepsilon_{\bf k}^2+\Delta_{\bf k}^2}}\right),
\end{equation}
where the gap $\Delta_{\bf k}$ can have $d$-wave symmetry of the
form $\Delta_{\bf k} = \Delta\cos(2\phi)$. Here $\Delta$ is the
gap amplitude and $\phi$ an angle along the cylindrical Fermi
surface. To keep things simple, we start with the $s$-wave case
and obtain
\begin{subequations}
\begin{equation}
  \label{eq:12}
  W^S_{KE}-W^N_{KE} = \frac{4\hbar^2}{m}\frac{N(0)}{D}
  \Delta^2\left[\ln\left(\frac{D}{\Delta}
  \right)-\frac{1}{2}\right].
\end{equation}
Eq.~\eqref{eq:12}
is the difference in KE between superconducting (S) and normal (N)
state which has increased as expected since $n_{\bf k}$ given by
Eq.~\eqref{eq:11a}
populates states above the chemical potential while the step function
of the normal state
does not. We have also worked out the difference in OS to give
\begin{equation}
  \label{eq:13}
  \bar{W}^S -\bar{W}^N = -\frac{N(0)\hbar^2}{m}
  \frac{2\Delta^2}{D},
\end{equation}
\end{subequations}
which has dropped in the superconducting state.
Knigavko {\it et al.}\cite{ref49} argued that $\bar{W}$ is the
quantity to use when discussing the OS. Here we note that $W$
itself shows no change with superconducting transition. Taking ratios
with the normal state in Eqs.~\eqref{eq:12} and \eqref{eq:13} we obtain
\begin{subequations}
  \begin{equation}
  \label{eq:13a}
  \frac{W^S_{KE}-W^N_{KE}}{\left\vert W^N_{KE}\right\vert} =
  \left(\frac{2\Delta}{D}\right)^2\left[\ln\left(\frac{D}
  {\Delta}\right)-\frac{1}{2}\right],
\end{equation}
and
\begin{equation}
  \label{eq:13b}
  \frac{\bar{W}^S-\bar{W}^N}{\bar{W}^N} =
  -\frac{1}{2}\left(\frac{2\Delta}{D}\right)^2.
\end{equation}
\end{subequations}
Note that because $\ln\left(D/\Delta\right)$ is expected to be
greater than one, the normalized KE change is greater than the value of the
OS drop in this very simplified model. So far we have considered only
$s$-wave. The formulas given above hold for $d$-wave with
$\Delta\to\Delta\cos(2\phi)$ and an average over angles is required.
When this is done
\begin{subequations}
\begin{equation}
\label{eq:13c}
\frac{W^S_{KE}-W^N_{KE}}{\left\vert W^N_{KE}\right\vert} =
\left(\frac{2\Delta}{D}\right)^2\frac{1}{2}\left[\ln\left(\frac{2D}
{\Delta}\right)-\frac{1}{2}\right],
\end{equation}
and
\begin{equation}
\label{eq:13d}
\frac{\bar{W}^S-\bar{W}^N}{\bar{W}^N} =
-\frac{1}{4}\left(\frac{2\Delta}{D}\right)^2.
\end{equation}
\end{subequations}
These are useful expressions to understand qualitatively the physics
but are not believed to be accurate. They do represent an extreme
case where the OS does not follow the temperature dependence of the
KE and they differ in the superconducting case as well.

\subsection{The Nearly Antiferromagnetic Fermi Liquid Model}
\label{ssec:2b}

So far we have not considered interactions yet the oxides are highly
correlated systems. A phenomenological approach to correlations
is embodied in the Nearly Antiferromagnetic Fermi Liquid model (NAFFL)
of Pines and coworkers\cite{ref35,ref36,ref37} and this approach is
convenient to obtain some information on the effect of interactions.
In this approach the important interactions proceed through the
exchange of spin fluctuations and the imaginary part of the spin
susceptibility enters a generalized set of Eliashberg equations.
A model susceptibility often used is\cite{ref37}
\begin{equation}
  \label{eq:5}
  \Im{\rm m}\left\{\chi({\bf q},\omega)\right\}
   = \frac{(\omega/\omega_{SF})\chi_{\bf Q}}
  {\left[1+\zeta^2({\bf q}-{\bf Q})^2\right]^2+(\omega/\omega_{SF})^2}.
\end{equation}
The parameters are $\chi_{\bf Q}$ the static susceptibility,
{\bf Q} the commensurate antiferromagnetic wave vector $(\pi/a,\pi/a)$
in the upper right hand quadrant of the first Brillouin zone and
symmetry related points. $\zeta$ is the magnetic coherence length
set at $\zeta = 2.5\,a$ in this paper and $\omega_{SF}$ is a
characteristic spin fluctuation frequency. Finally, $g$ is the
coupling between charge carriers and the spin fluctuations.
The Eliashberg equations for renormalized Matsubara frequencies
$\tilde{\omega}({\bf k},i\omega_n)$, renormalized energies
$\xi({\bf k},i\omega_n)$, and pairing energy (gap) in the
\begin{figure}[tp]
\vspace*{-5mm}
  \includegraphics[width=9cm]{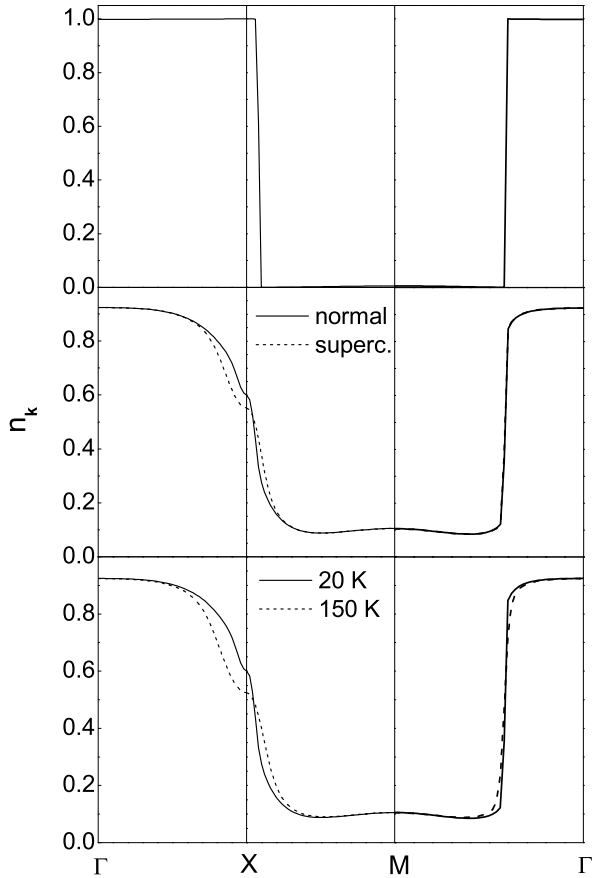}
  \caption{The occupation number $n_{\bf k}$ for
selected directions in the CuO$_2$ Brillouin zone. Model A of
Table~\protect{\ref{tab:1}} was used.
Top frame: the non interacting case at $T=0$. It is included for
comparison. Center frame: The interacting
case at a temperature $T=20\,$K. We show normal state (solid line)
and superconducting state (dashed line) results. Bottom frame: The
temperature influence on the normal state $n_{\bf k}$ for
$T=20\,$K (solid line) and $T=150\,$K (dashed line). }
  \label{fig:2}
\end{figure}
superconducting state
 $\phi({\bf k},i\omega_n)$
are\cite{ref35,ref38,ref39,ref40,ref40a}
\begin{widetext}
\begin{subequations}
\label{eq:6}
\begin{eqnarray}
  \tilde{\omega}({\bf k},i\omega_n) &=& \omega_n+
  T\sum\limits_m\sum\limits_{{\bf k}'}
  \lambda_{SF}({\bf k}-{\bf k}',i\omega_n-i\omega_m)
  \frac{\tilde{\omega}({\bf k}',i\omega_m)}{D({\bf k}',i\omega_m)},
  \label{eq:6a}\\
  \xi({\bf k},i\omega_n) &=&
  -T\sum\limits_m\sum\limits_{{\bf k}'}
  \lambda_{SF}({\bf k}-{\bf k}',i\omega_n-i\omega_m)
  \frac{\epsilon_{{\bf k}'}+\xi({\bf k}',i\omega_m)}{D({\bf k}',i\omega_m)},
  \label{eq:6b}\\
  \phi({\bf k},i\omega_n) &=&
  -T\sum\limits_m\sum\limits_{{\bf k}'}
  \lambda_{SF}({\bf k}-{\bf k}',i\omega_n-i\omega_m)
  \frac{\phi({\bf k}',i\omega_m)}{D({\bf k}',i\omega_m)},
  \label{eq:6c}
\end{eqnarray}
with
\begin{equation}
  \label{eq:6d}
  D({\bf k},i\omega_n) = \tilde{\omega}^2({\bf k},i\omega_n)+
  \left[\epsilon_{\bf k}+\xi({\bf k},i\omega_n)\right]^2+
  \phi^2({\bf k},i\omega_n),
\end{equation}
\end{subequations}
and
\begin{equation}
  \label{eq:7}
  \lambda_{SF}({\bf q},i\nu_n) = \frac{g^2\chi_{\bf Q}}
  {1+\zeta^2({\bf q}-{\bf Q})^2+(\vert\nu_n\vert/\omega_{SF})}.
\end{equation}
Here {\bf q} is the momentum transfer and
$i\omega_n$, $i\nu_n$ are the electron and boson Matsubara frequencies,
respectively. Finally, the filling $\langle n \rangle$ is determined
from
\begin{equation}
  \label{eq:8}
  \langle n\rangle = \frac{1}{2}-\sum\limits_{\bf k}\sum\limits_{n\ge 0}
  \frac{\epsilon_{\bf k}+\xi({\bf k},i\omega_n)}
  {\tilde{\omega}^2({\bf k},i\omega_n)+\left[\epsilon_{\bf k}+
  \xi({\bf k},i\omega_n)\right]^2+\phi^2({\bf k},i\omega_n)}.
\end{equation}
\end{widetext}
For fixed filling $\langle n\rangle$ this last equation determines the
chemical potential which changes with temperature for bands which do
not have particle-hole-symmetry.

We solved the Eliashberg
equations\cite{ref40} for the two models of Table~\ref{tab:1} with
the single parameter not yet specified, $g$, adjusted to get a
critical temperature of $90\,$K and $100\,$K, respectively, for model A
and model B. The results for $W$ (solid down-triangles) and
$-W_{KE}/2$ (solid up-triangles) are shown in Fig.~\ref{fig:1} in which
they are compared with the non interacting results. [A $128\times 128$
sampling of the \textbf{k}-space, $ak_x$, $ak_y\in[0,\pi]$ was used
which accounts for the wiggles. In one case (not shown in the figure)
the \textbf{k}-space sampling was increased to $512\times 512$. This
made the curves smoother but there were no other changes.]
In both models
the interactions have lowered the value of each integral but the
temperature dependence of $W$ and $-W_{KE}/2$ still largely
track each other. The role played by the interactions can be
traced simply in Eqs.~\eqref{eq:1} and \eqref{eq:3} as changing the
probability of occupation of the state $\vert{\bf k},\sigma\rangle$
factor $n_{{\bf k},\sigma}$.
This factor depends on correlations as seen in Fig.~\ref{fig:2}
where we show results for model A. The top frame gives $n_{\bf k}$
(for both spins)
along $\Gamma\to X\to M\to \Gamma$ in the first Brillouin zone for the
non interacting case while the bottom frame shows results at $T=20\,$K
(solid curve) and $T=150\,$K (dotted curve) in the normal state. We
see that the effect of interactions is to make the probability of
occupation of the state $\vert{\bf k}\rangle$
non zero even in regions where there
is no occupation in the non interacting case. Also, there is
considerable ``smearing'' of the interacting case distribution which
changes with temperature and with the onset of superconductivity (center
frame). On comparing bottom and center frame we note an increase
in temperature of about $100\,$K in the normal state
corresponds roughly to the same amount
of extra smearing as is due to the onset of superconductivity. This
smearing in the occupation factor means an increase in kinetic
energy. This is expected in a BCS mechanism. In the Cooper pair
model two electrons are introduced at the Fermi surface of a
quiescent Fermi sphere. Because of an effective attractive potential
between them they prefer to go into a superposition of plane wave
states with $\vert{\bf k}\vert$ equal or slightly greater than
$k_F$, the Fermi momentum,
and so reduce their potential energy. Although the kinetic
energy in the process is increased over the free electron value
of $2\varepsilon(k_F)$, the potential is sufficiently reduced to
compensate for this increase.

Returning to Fig.~\ref{fig:1} several features are to be noted.
For the case of no interactions the numerical data for both
$W$ and $W_{KE}$ is well fit by a $T^2$ law as is also
the interacting case based on Model A. But this is clearly not
so for Model B for which the OS and also $-W_{KE}/2$ turns up
from a $T^2$ law as $T$ is lowered towards zero with turn up onset
occurring around $T=100\,$K. In this case the change in $W$
between $T=0$ and $200\,$K is 6.4\% which is much larger than for
the non interacting case (1.5\%). While the two models A and B have a
different Fermi surface, dispersion relation, and filling, the
most important parameter
\begin{figure}[tp]
\vspace*{-3mm}
%\centering
  \includegraphics[width=9cm]{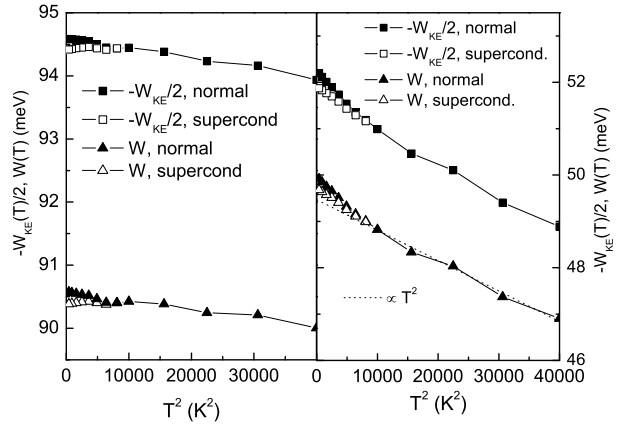}
 \vspace*{-5mm}
  \caption{Comparison of normal and superconducting state for the
optical sum and the kinetic energy. The left hand frame applies to
Model A of Table~\ref{tab:1} and $\omega_{SF}=82\,$meV and the
right hand frame is for the band structure Model B of Table~\ref{tab:1} and
$\omega_{SF} = 10\,$meV. The dotted line in this frame shows
a $T^2$ law extrapolation of the normal state data for $T>T_c$ to
zero temperature. Adapted from Ref.~\protect{\onlinecite{ref40}}.
}
  \label{fig:3}
%\vspace*{5cm}
\end{figure}
which gives the deviation from the $T^2$ law noted above is the small
value of the spin fluctuation frequency $\omega_{SF}=10\,$meV
used in Model B. This implies that the thermally activated scattering is
larger in this model than it is for Model A. Later, when we consider
simpler versions of our interaction model, we will return to this
issue where it will become much easier to trace these dependencies.

In Fig.~\ref{fig:3} we compare normal and superconducting state for the
two models of Table~\ref{tab:1}. We see that both KE and OS (open squares
and up triangles, respectively) fall below their normal state
values (solid squares and up triangles, respectively) at the same
temperature. This corresponds to an increase in KE as the
superconducting state is entered. The changes are small in all
cases. The KE in meV has increased by 0.2\% for Model A and
by 0.77\% in Model B between superconducting and normal state at
$T=0$. For the OS the differences are 0.22\% and 0.6\%, respectively,
very comparable to what is found for the KE but not identical. We
note that in the NAFFL model once the susceptibility \eqref{eq:7}
is specified, superconducting solutions with $d$-wave symmetry result.
These are not put in by hand as to symmetry or functional form.
The gap involves a harmonic $[\cos(ak_x)-\cos(ak_y)]$ as well as
many of the higher harmonics consistent with the
$d_{x^2-y^2}$ irreducible representation of the symmetry group
for the square CuO$_2$ lattice. The solutions are far from simple
BCS. In $d$-wave BCS an ansatz on the pairing potential
$V_{{\bf k},{\bf k}'}$ is made that it have the form
$\eta_{\bf k}V \eta_{{\bf k}'}$ with $\eta_{\bf k}\sim[\cos(ak_x)-\cos(ak_y)]$.
This leads to a $d$-wave gap consisting of the lowest harmonic only.
Early solutions\cite{ref38}
based on Model A with $\omega_{SF} = 7.76\,$meV showed that the gap
had maximum amplitude at the $X$-point $(\pi,0)$
of the two dimensional CuO$_2$
Brillouin zone with $\Delta_{max}^{BZ} = 33\,$meV. On the other
hand the maximum gap on the Fermi surface was $\Delta_{max}^{FS}=27\,$meV at
$T=20\,$K far below its BZ maximum. The value of
$2\Delta_{max}^{FS}/(k_BT_c) = 6.27$ is quite different from the BCS
value of 4.28. Also shown on the right hand panel is a dotted
straight line which indicates the least squares fit extrapolation of
the normal state data to zero temperature from $T>T_c$. It represents
a $T^2$ law. It is important to notice, for later reference, that both,
the superconducting as well as normal state $W$ data are above this
dotted line for all temperatures $T<T_c$.

\subsection{The Hubbard Model and Dynamical Mean Field}
\label{ssec:2c}

%Throughout this section we follow the notation of Toschi
%{\it et al.}\cite{ref42} and use the symbol $D$ for half the bandwidth
%in contrast to the main body of this review in which $D$ symbolizes the full
%band width.
%
Another approach used to treat strongly correlated systems such as
the cuprates is Dynamical Mean Field Theory (DMFT).\cite{ref41} The
approach is numerical and based on the Hubbard model with hopping
$t$, onsite Coulomb repulsion $U$, and chemical potential $\mu$.
The half bandwidth $D = 8\,t$ is taken to be $1.2\,$eV in what
follows with $U = 3D/2 = 12\,t$, so that the antiferromagnetic
super exchange $J = 4\,t^2/U\simeq 100\,$meV. For large values of
$4U/D$ at half filling the system is a Mott insulator. Away from
half filling, the weight of the quasiparticle peak at the Fermi
level denoted by $Z$ and related to the self energy $\Sigma(\omega)$
by 
\[
%      \label{eq:9}
   Z = \left[1-\left.\frac{\partial\Sigma(\omega)}{\partial \omega}
  \right\vert_{\omega=0}\right]^{-1}   
\]
is non zero but small and is a measure of the metalicity.

The Hamiltonian used has the form
\begin{equation}
  \label{eq:9}
  H = -t\sum\limits_{\langle ij\rangle\sigma}c^\dagger_{i\sigma}
  c_{j\sigma}+U\sum\limits_i n_{i\uparrow} n_{j\downarrow}-
  \mu\sum\limits_i\left(n_{i\uparrow}+n_{i\downarrow}\right),
\end{equation}
where $\sigma$ is spin $\uparrow$, $\downarrow$, $c_{i\sigma}$
($c^\dagger_{i\sigma}$) are annihilation (creation) operators
for fermions of spin $\sigma$ on site $i$, $n_{i\sigma} =%
c^\dagger_{i\sigma}c_{i\sigma}$, and the sum $\langle ij\rangle$
\begin{figure}[tp]
%  \vspace*{0.5cm}
%  \centering
  \includegraphics[width=8.3cm]{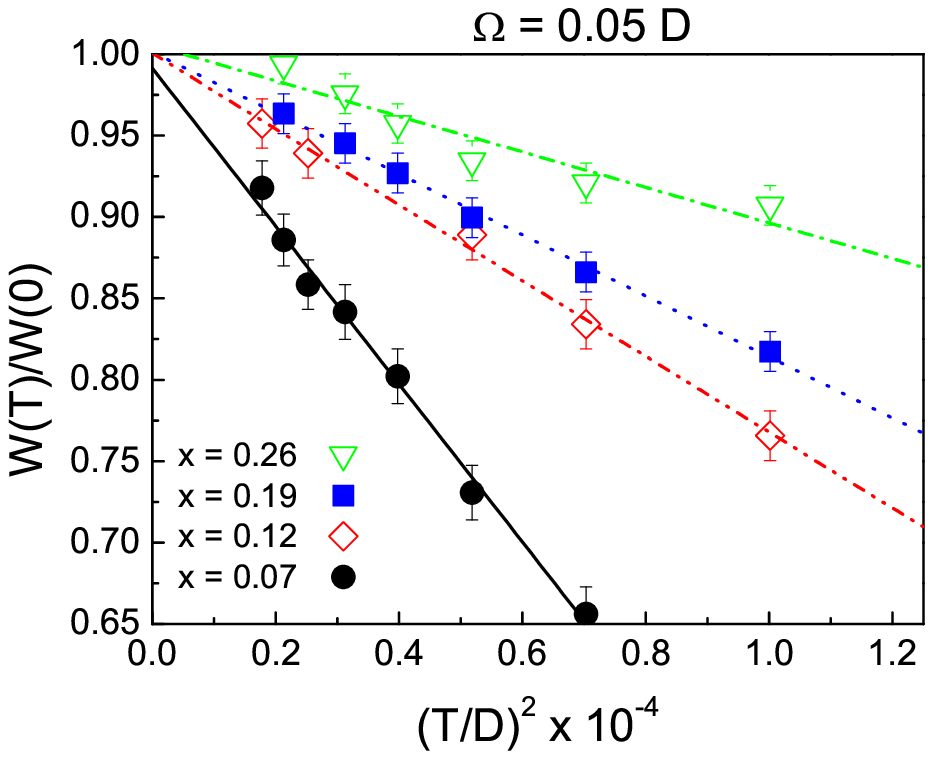}
  \includegraphics[width=7.9cm]{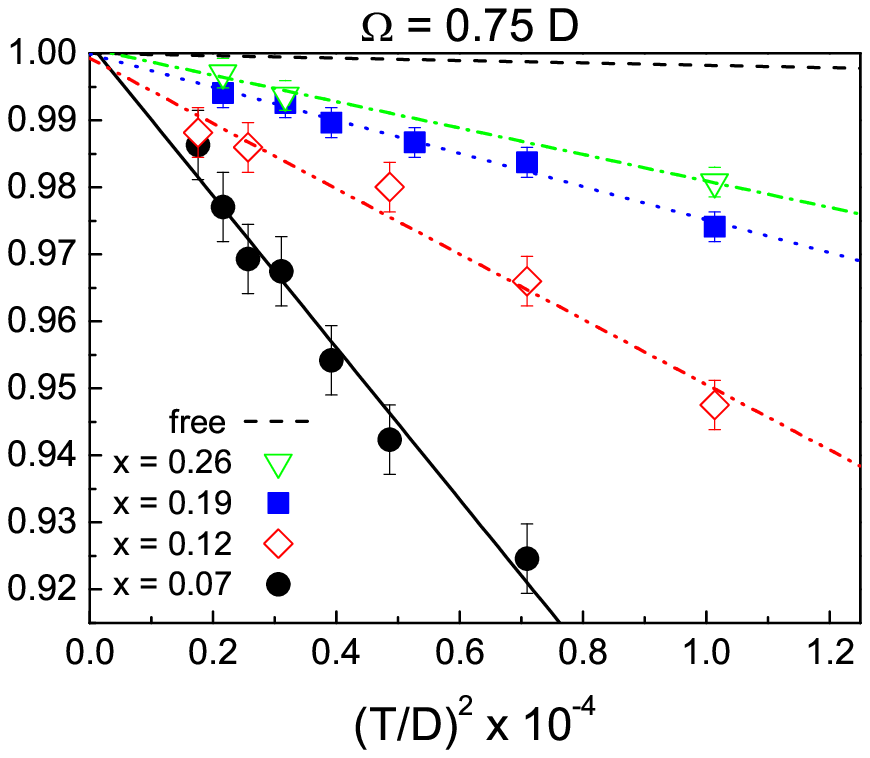}
  \caption{(Color online) $W(\Omega,T)$ normalized to its $T=0$ value
as a function of $T^2$ for the Hubbard model,
for $\Omega = 0.05\,D$ (top frame) and $\Omega = 0.75\,D$
(bottom frame).
Symbols are the results of the DMFT calculations; lines are best
fits to them. The various symbols give four dopings $(x)$.
Adapted from Ref.~\protect{\onlinecite{ref42}}.
}
  \label{fig:4}
\end{figure}
is restricted to nearest neighbors only. In DMFT the full many body
system with strong interactions is modeled as a single site
problem with relevant interactions between, say, two electrons at
that site plus coupling to a bath which represents on
average the remaining degrees of freedom. The bath and local
problem is to be solved for in a self consistent way. The method is
now well developed and has proved its ability to simultaneously
describe low and high energy features in Mott systems.
Toschi {\it et al.}\cite{ref42} have calculated the real part
of the optical conductivity in this model and obtained the results
\begin{figure}[tp]
%  \vspace*{5mm}
  \includegraphics[width=8.5cm]{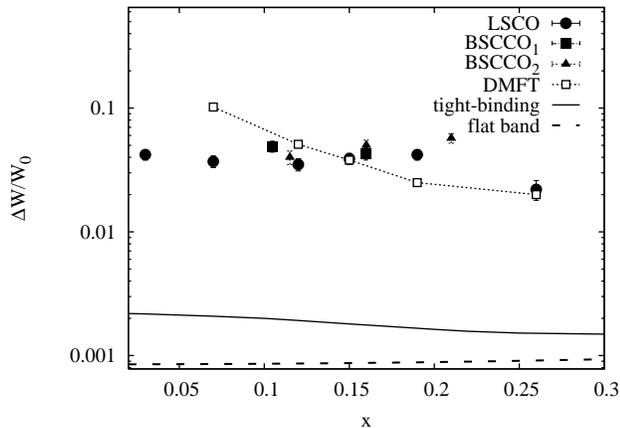}
  \caption{The relative variation of spectral weight between the
lowest $T$ and $300\,$K as a function of doping $x$ for various
cuprates (full symbols) is compared with DMFT calculations
(open squares) and with the predictions of non interacting models
[tight binding model (solid line), constant density of states
approximation (dashed line)]. The
dotted line is a guide to the eye. The simple inclusion of
correlation effects allows one to reproduce the observed absolute values
with no need of fitting parameters. Data for LSCO are obtained from
Refs.~\protect{\onlinecite{ref43}} and \protect{\onlinecite{ref44}},
and for BSCCO from
Refs.~\protect{\onlinecite{ref45,ref46,ref47}}
(BSCCO$_1$) and from Ref.~\protect{\onlinecite{ref48}} (BSCCO$_2$).
Adapted from Ref.~\protect{\onlinecite{ref42}}.
}
  \label{fig:5}
\end{figure}
\begin{figure}[tp]
  \centering
  \includegraphics[width=9cm]{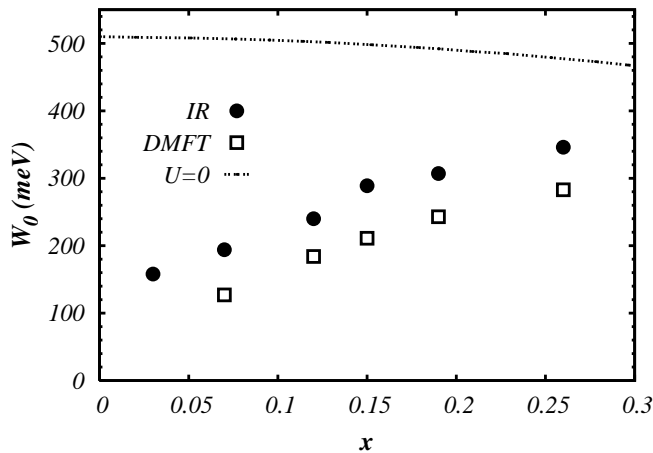}
  \caption{The $T=0$ spectral weight $W_0$ for $\Omega = 0.75D$ as a
function of doping $x$ for the Hubbard model (open squares) and the
experimental observation of Lucarelli {\it et al.}
Ref.~\protect{\onlinecite{ref44}} (solid circles) obtained by integrating
data on $\sigma_1(\omega)$ for LSCO. The dotted curve is for reference
and gives results for $U=0$ (no correlations included). Adapted by
A. Toschi from Ref.~\protect{\onlinecite{ref42}}.
}
  \label{fig:4a}
\end{figure}
shown in Fig.~\ref{fig:4} for various values of the doping $x$ as
indicated in the figure which gives $W(\Omega,T)/W(\Omega,T=0)$
as a function of $(2T/D)^2$. The left hand frame employs a cutoff
$\Omega$ in Eq.~\eqref{eq:1} of $0.05D$ while the right hand
frame is for $\Omega = 0.75D$. All curves appear to follow a $T^2$
behavior and the slope of
these lines becomes smaller with increasing $\Omega$.
A comparison of their DMFT results with experiment is also
offered by the same authors and we reproduce it in Fig.~\ref{fig:5}
where we show
the difference $\Delta W \equiv W(T=0)-W(T=300\,{\rm K})$ for
$\Omega=1.5\,D$ renormalized to $W_0=W(T=0)$ denoted by $\Delta W/W_0$
on the figure as a function of doping $x$ (open squares).

Experimental
results are for La$_{2-x}$Sr$_x$CuO$_4$ (LSCO, solid circles), Refs.~%
\onlinecite{ref43} and \onlinecite{ref44},
and Bi$_2$Sr$_2$CaCu$_2$O$_{8+x}$. Solid squares
are from Refs.~\onlinecite{ref45,ref46,ref47} (BSCCO$_1$) and solid
triangles from Ref.~\onlinecite{ref48} (BSCCO$_2$). Also shown for
comparison are results for tight binding with no interactions (solid line)
and in a constant density of states approximation (dashed line).
It is clear that
correlations dominate the observed change in $W$ between zero and
$300\,$K to its $T=0$ value.
DMFT provides a reasonable fit to the data while tight binding fails badly,
giving values that are too small.

In Fig.~\ref{fig:4a} DMFT results (open squares) as a function of
doping $x$ are compared with the experimental results (sold circles)
of Lucarelli {\it et al.}\cite{ref44} for the optical sum
of LSCO at $T=0$.
The theory reproduces well the doping dependence of
$W_0\equiv W(\Omega=0.75D,0)$. While the absolute value of $W_0$ is
somewhat underestimated its large decrease as the Mott transition
is approached is well captured by the model calculations. Also shown
on the same figure (dotted line) are results obtained in the limit
$U=0$ (no Hubbard $U$). We see that in this case the observed trend
with doping is not reproduced and $W_0$ is much too large, particularly
at small dopings indicating, as expected, that near the Mott transition
$W_0$ goes like $x$ rather than $1-x$. Finally, we note that the
inclusion of $U$ (correlation effects) reduces $W_0$ just as we have
seen in the NAFFL model. We note, however, that this model is not well
suited to describe doping differences because the phenomenological
spin susceptibility \eqref{eq:5} really needs to be fixed to some
experimental observation at each doping level, particularly as the
Mott transition is approached. 

One thing to note about the DMFT results shown in Fig.~\ref{fig:4} is
that even for the smaller value of the cutoff $\Omega$ on the OS, a
$T^2$ behavior is observed at least for the limited data available
in the figure. By contrast, we have seen in the NAFFL calculations
of Figs.~\ref{fig:2} and \ref{fig:3} that deviations from $T^2$ can
occur when the characteristic spin fluctuation energy $\omega_{SF}$
is small even when the sum is taken over the entire band, i.e.:
$\Omega$ is large enough to include all contributions. Thus, the
two calculations differ in this important point. We will see later
that there is no guarantee that a given model for the interactions
should always give a $T^2$ law.

To end we mention related Hubbard
model work by Maier {\it et al.}\cite{ref89c} They work directly
with the KE rather than with the conductivity as did Toschi
{\it et al.}\cite{ref42} and use the dynamical cluster approximation
(DCA). They present results in both normal and superconducting
state for two doping levels, namely $x = 0.05$ and
$x=0.2$. In both cases the KE is found to decrease with
decreasing temperature even in the superconducting state and this
decrease, in fact, drives superconductivity. The changes are of order
a few percent at most.

As we have been so far mainly interested in
the normal state we postpone further discussion of this work to
a later section which deals explicitly with the superconducting state.
We will also describe the recent work of Haule and Kotliar\cite{ref89b}
which deals with the $t$-$J$ model (see Sec.~\ref{sec:4}).

\section{The effect of interactions in isotropic boson
exchange models}
\label{sec:3}
\subsection{Self Energy in Finite Bands}
\label{ssec:3b}

Recent studies of finite band effects have revealed that the self
energy due to impurities or to interaction with bosons is profoundly
changed by the application of a cutoff $\pm D/2$ in a half filled
band.\cite{ref54,ref55,ref56,ref57,ref58,ref59,ref62}
In a finite band, the normal state self energy due to the
electron-phonon (or some other boson) interaction is given by
\begin{subequations}
\label{eq:23s}
\begin{eqnarray}
  \Sigma(z) &=& T\sum\limits_m\lambda(z-i\omega_m)\eta(i\omega_m)\nonumber\\
  &&+
  \int\limits_0^\infty\!d\nu\,\alpha^2F(z)\left\{
  [f(T,\nu-z)+n(T,\nu)]\right.\nonumber\\
  &&\left.\times\eta(z-\nu)+[f(T,\nu+z)+n(T,\nu)]\eta(z+\nu)\right\},
  \nonumber\\
  \label{eq:23}
\end{eqnarray}
in the mixed representation of Marsiglio {\it et al.}\cite{ref60}
Here,
\begin{equation}
  \label{eq:24}
  \lambda(z) = \int\limits_0^\infty\!d\nu\alpha^2F(\nu)\frac{2\nu}
   {\nu^2-z^2}
\end{equation}
and
\begin{equation}
  \label{eq:25}
  \eta(z) = \int\limits_{-\infty}^\infty\!d\epsilon\,
  \frac{N_0(\epsilon)}{N_0(0)}\frac{1}{z-\epsilon-\Sigma(z)}.
\end{equation}
\end{subequations}
These equations need to be solved self consistently
as $\Sigma(z)$ itself depends on itself through Eq.~\eqref{eq:25}.
Here $z$ is a complex variable, $n(T,\nu)$ is the Bose thermal factor,
$\alpha^2F(z)$ the electron-phonon spectral density and
$N_0(\epsilon)/N_0(0)$ the normalized bare electronic density of
states taken here to be constant and confined to $[-D/2,D/2]$.
This provides a band cutoff in $\eta(z)$ of Eq.~\eqref{eq:25}
and makes Eq.~\eqref{eq:23} self consistent. These equations are also
given by Karakozov and Maksimov\cite{ref53} where they are written
either in pure Matsubara notation or fully on the real frequency
axis rather than in the mixed notation of Eqs.~\eqref{eq:23s} %to
%\eqref{eq:25} 
which includes both versions, Matsubara and real axis.
We have found the mixed notation more convenient for numerical
work. For coupling to a
single Einstein boson mode $\alpha^2F(\omega) = a\delta(\omega-\Omega_E)$
where $\Omega_E$ and $a$ are taken in units of $D/2$. The electron mass
renormalization is given by $\lambda = 2a/\Omega_E$ which is
dimensionless. For an
infinite band at $T=0$ we recover from Eqs.~\eqref{eq:23s} %to
%\eqref{eq:25} 
the very familiar result\cite{ref62}
\begin{equation}
  \label{eq:26}
  \Sigma_1(\omega) = a\ln\left\vert\frac{\omega-\Omega_E}{\omega+
  \Omega_E}\right\vert,
\end{equation}
and
\begin{equation}
  \label{eq:27}
  \Sigma_2(\omega) = -\pi a \theta\left(\vert\omega\vert-\Omega_E\right),  
\end{equation}
where $\theta(x)$ is the theta function equal to one for $x>0$ and
zero otherwise. For positive values of $\omega$ the real part
$\Sigma_1(\omega)\le 0$ and approaches  zero as $\omega\gg\Omega_E$.
The imaginary part $\Sigma_2(\omega)$ is zero till
$\vert\omega\vert = \Omega_E$ at which point it jumps to a value
of $\pi a$ and then stays constant. Finite band effects change this
behavior radically.

\begin{figure}[tp]
%  \centering
  \includegraphics[width=9cm]{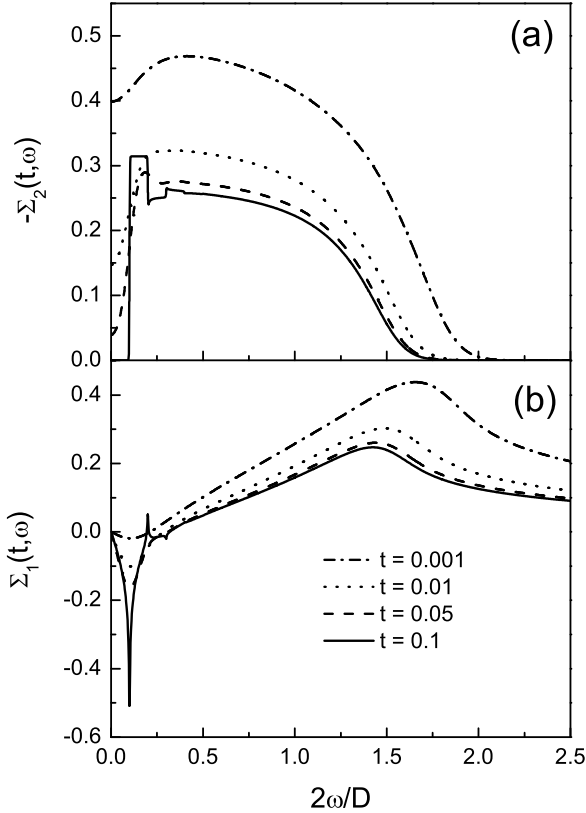}
  \caption{Evolution of the self energy vs frequency dependence with
temperature for the Einstein spectrum with $a=0.1$ and $\Omega_E =
0.1$ $(\lambda = 2)$. Frame (a) is for minus the imaginary part,
$-\Sigma_2(\omega)$, while frame (b) is for the real part,
$\Sigma_1(\omega)$. In each frame the different curves correspond
to temperatures $t\equiv 2T/D =0.1$ (dash-dotted line),
0.05 (dotted line),
0.01 (dashed line), 0.001. All energies are in units of $D/2$.
Adapted from Ref.~\protect{\onlinecite{ref58}}.
}
  \label{fig:7}
\end{figure}

In Fig.~\ref{fig:7} we show results of Knigavko and Carbotte\cite{ref58}
for minus the imaginary part $[-\Sigma_2(\omega)]$ of the self energy
in frame (a) and its real part $[\Sigma_1(\omega)]$ in frame (b) for
%\clearpage
coupling to a single Einstein phonon at $\Omega_E = 0.1$ with
mass enhancement parameter $\lambda = 2$ for different temperatures,
all in units of $D/2$.
The temperatures are as given in the figure caption.
The sharper curves correspond to the lower temperatures.
 For $T=0$, $\Sigma_1(\omega)$ has a
logarithmic like singularity at $\omega = \Omega_E$ as in the
infinite band case of Eq.~\eqref{eq:26} but now, rather than remain
negative as it goes towards zero for $\omega\gg \Omega_E$,
$\Sigma_1(\omega)$ crosses the $\omega$-axis and takes on large
positive values before dropping towards zero from above. Equally
different from the infinite band case, $-\Sigma_2(\omega)$ is not
a constant equal to $\pi a$ [see Eq.~\eqref{eq:27}] above
$\omega>\Omega_E$ but rather drops as we approach the bare band
edge at $\omega=1.0$ after which it becomes small for
$\omega\stackrel{>}{\sim}1.5$ which is where the renormalized
density of states $\tilde{N}(\omega)$ also becomes small. For further
discussion of finite band features we refer the reader to Refs.
\onlinecite{ref54} to \onlinecite{ref62} as well as to Ref.~%
\onlinecite{ref53} where a somewhat different density of states
model is used but this does not change the qualitative behavior of the
self energy seen in Fig.~\ref{fig:7}. An advantage of Karakozov and
Maksimov's  choice of
DOS, however, is that they can get a simple analytic expression
for the self energy which they also show to be accurate by comparing
with numerical results based on the full Eliashberg equations.

We return next to the OS of Eq.~\eqref{eq:17}. % which can be rewritten
%as:
%\begin{equation}
%  \label{eq:28}
%  \bar{W}(T) = \frac{\hbar^2}{2m_b}\left[1-\frac{4}{(D/2)^2}
%  \int\limits_0^{D/2}\!d\epsilon\,\epsilon\int\limits_{-\infty}^\infty\!
%  d\omega\,f(T,\omega)A(\epsilon,\omega)\right].
%\end{equation}
One can get insight into the relative importance of the temperature
dependence carried by the thermal factor $f(T,\omega)$ in
$\bar{W}(T)$ and the temperature
\begin{figure}[tp]
%  \centering
  \includegraphics[width=9cm]{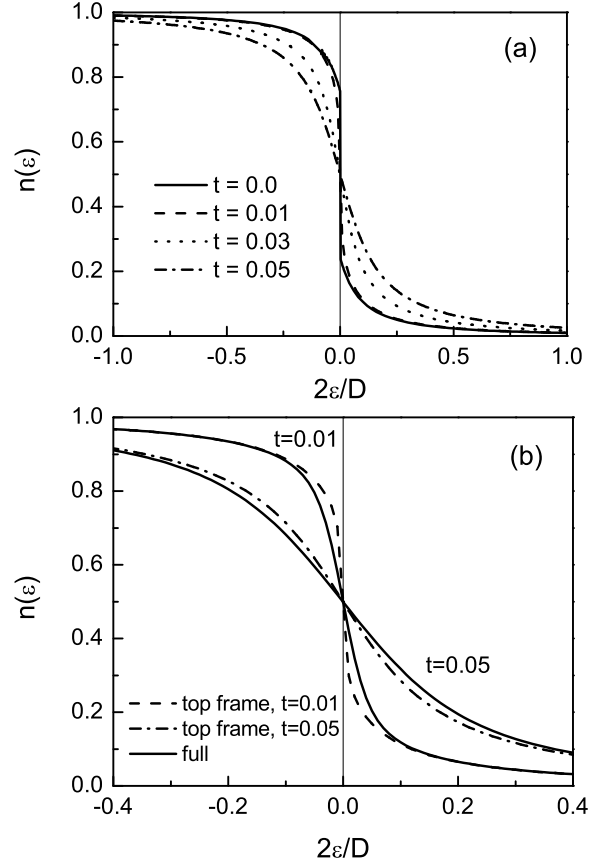}
  \caption{The probability of occupation $n(\epsilon)$ of a state
of energy $\epsilon$ vs
normalized energy $2\epsilon/D$. (a) The results for the case when only
the temperature dependence of the self energy is included. The normalized
temperatures are $t \equiv 2T/D = 0.0$ (solid line), 
0.01 (dashed line), 0.03 (dotted line), and 0.05
(dash-dotted line). (b) Comparison of the
results when the complete temperature dependence  of $n(\epsilon)$
[see Eq.~\protect{\eqref{eq:17}}] is accounted for (solid lines)
with the case presented in frame (a). The normalized
temperatures are $t=0.01$ (dashed line) and 0.05 (dash-dotted
line). (All energies are in units of $D/2$.) A constant density of states
model with cutoff $\pm D/2$ is used. Adapted from
Ref.~\protect{\onlinecite{ref49}}.
}
  \label{fig:8}
\end{figure}
dependence solely due
to the interactions in $A(\epsilon,\omega)$. In Fig.~\ref{fig:8}
we show results obtained by Knigavko {\it et al.}\cite{ref49}
for $n(\epsilon)$, Eq.~\eqref{eq:10}.
Frame (a) shows $n(\epsilon)$
vs the renormalized energy $2\epsilon/D$ for four values of reduced
temperatures $t=2T/D$,
namely 0.0 (solid line), 0.01 (dashed line),
0.03 (dotted line), and 0.05 (dash-dotted line)
when only the temperature dependence
in $A(\epsilon,\omega)$ is included, i.e.: the $f(T,\omega)$ factor
in $n(\epsilon)$ was excluded. Here we have considered coupling to a
single phonon of energy $\Omega_E=0.04$ with mass enhancement
$\lambda = 1$.
It is clear that considerable temperature dependence comes from this
source alone
and will influence the temperature dependence of the
OS. In frame (b) of Fig.~\ref{fig:8} we show results
(solid curves)
when both sources of $T$ dependence in $n(\epsilon)$ are included
and compare with the two equivalent curves of frame (a) for
$t=0.01$ (dashed line) and 0.05 (dash-dotted line).
It is clear from these figures that the temperature
dependence of the self energy is always important in determining
the probability of occupation, $n(\epsilon)$, of a state of energy
$\epsilon$ and hence the
OS's $T$ dependence.

\subsection{Results for a delta-function boson model}
\label{ssec:3c}

\begin{figure}[tp]
%  \centering
\vspace*{-7mm}
  \includegraphics[width=9cm]{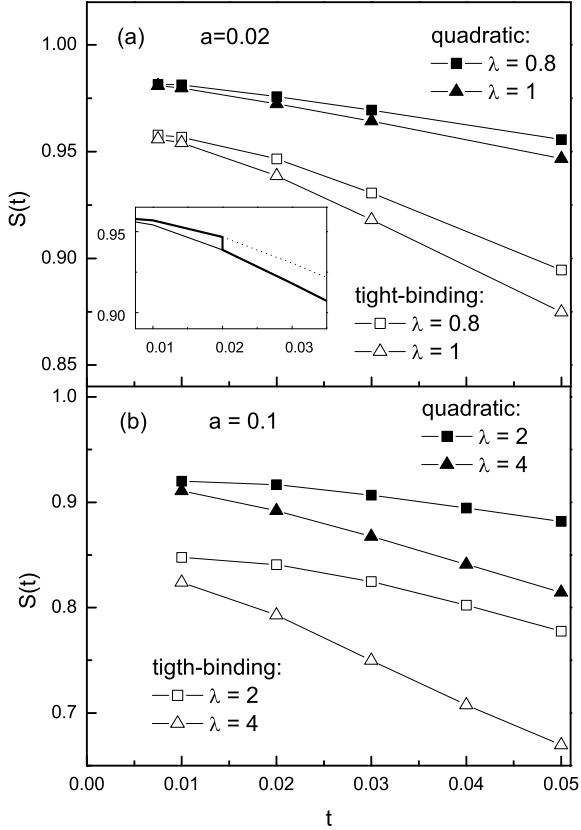}
  \caption{The variation of the optical sum $S$ vs reduced temperature
$t = 2T/D$ for the
interaction strengths (a) $a=0.02$ and (b) $a=0.1$. The results are for
both the quadratic (solid symbols) and tight binding
(open symbols) bands, as indicated. In the top frame
the mass enhancement factor $\lambda=0.8$ (squares) and
$\lambda=1$ (triangles), while in the bottom frame $\lambda=2$ (squares)
and $\lambda=4$ (triangles).
The inset in the top frame illustrates
the behavior of the optical sum during a sudden ``undressing
transition'' at $t_{undress}=0.02$ with 20\% hardening of the
normalized boson frequency $\Omega$ and a corresponding reduction
of the mass enhancement factor from $\lambda=1.0$ to 0.8. Adapted
from Ref.~\protect{\onlinecite{ref49}}.
}
  \label{fig:9}
\vspace*{-3mm}
\end{figure}
 Results for the OS normalized by a factor
of $\hbar^2/(2m_b)$ for the tight binding case \eqref{eq:17}
and by $\hbar^2\, n/m$ for the quadratic dispersion, both
denoted by $S(t)$, are given in Fig.~\ref{fig:9} reproduced
from Knigavko {\it et al.}\cite{ref49} The parameters are as shown on the
figure and given in the caption. The top frame is for $a=0.02$ and
the bottom frame for $a=0.1$. The former corresponds
to a conventional metal while the latter may be more characteristic of
the cuprates. Two representative values of the boson energy
$\Omega_E$ have been chosen to get $\lambda = 0.8$
(squares) and 1.0 (triangles)
in the top frame and $\lambda = 2$ (squares) and
$\lambda = 4$ (triangles) in the bottom frame.
Both tight binding (open symbols)
and quadratic (solid symbols) dispersion
relations are considered. It is quite clear from
these results that the temperature dependence of the OS needs not be
quadratic and depends on the size of the coupling to the phonons
($\lambda$ value) and also on the dispersion relation used to describe
the electron dynamics. An important observation for what will follow
is contained in the inset of the top frame of Fig.~\ref{fig:9}. The
heavy solid line follows the temperature evolution of the OS when
a sharp ``undressing'' transition is assumed to take place at $t=0.02$
where a 20\% hardening of the phonon spectrum occurs.
This corresponds to a shift in the mass enhancement parameter from
$\lambda=1$ to 0.8. Such a hardening of the phonon energy leads to
an increase in the OS corresponding to a decrease in the KE of the
charge carriers.\cite{ref49}

\subsection{Results for an MMP spin fluctuation model}
\label{ssec:3d}

The delta function spectra used in the work of Knigavko {\it et al.}%
\cite{ref49} while giving us insight
into the mechanism that leads to temperature dependences in the OS is not
realistic for the cuprates. If one wishes to remain within the framework of
a boson exchange mechanism one should at the very least choose a different
form for $\alpha^2F(\omega)$. For spin fluctuation exchange a much
used choice is the MMP model of Millis {\it et al.}\cite{ref37,ref61,ref63}
In its simplest form $\alpha^2F(\omega)$ which will now be denoted
$I^2\chi(\omega)$ is taken as
\begin{equation}
  \label{eq:29}
  I^2\chi(\omega) = I^2\frac{\omega/\omega_{SF}}{1+(\omega/\omega_{SF})^2},
\end{equation}
with $I$ a coupling between charge carriers and spin susceptibility
and $\omega_{SF}$ a spin fluctuation frequency. Here, the coupling $I$ is
adjusted to get a superconducting transition temperature $T_c$ of
\begin{figure}[tp]
%  \centering
  \includegraphics[width=9cm]{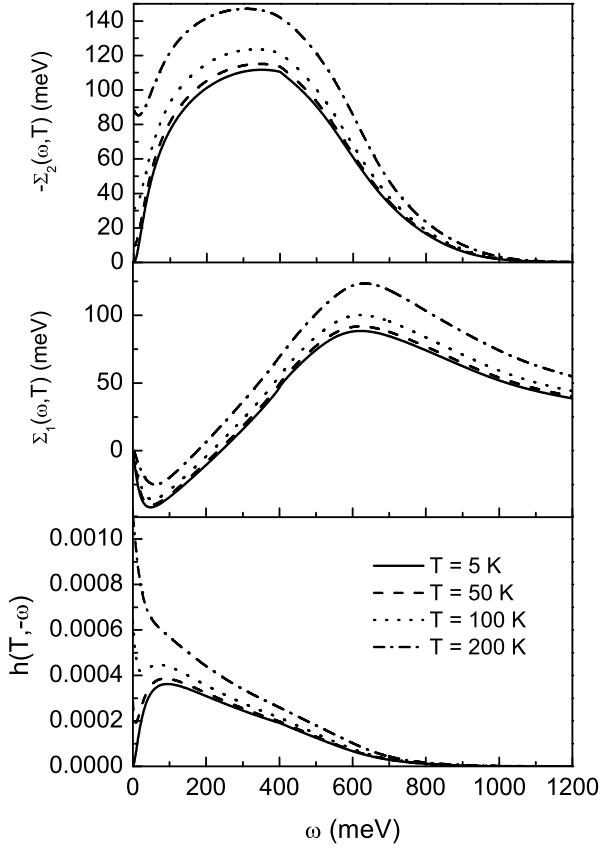}
  \caption{Top frame: minus the imaginary part of the carrier self
energy $-\Sigma_2(\omega,T)$ vs $\omega$ at the various temperatures
shown. Middle frame: minus the real part of the carrier self energy
$\Sigma_1(\omega,T)$ vs $\omega$. Bottom frame: the function
$h(T,-\omega)$ of Eq.~\protect{\eqref{eq:18}} vs $\omega$. The
calculation is for an MMP model, Eq.~\protect{\eqref{eq:29}}, with
$\omega_{SF} = 20\,$meV, $I^2=0.82$, and $D=800\,$meV.
}
  \label{fig:10}
\end{figure}
$100\,$K when used in the corresponding Eliashberg
equations generalized for a $d$-wave gap symmetry. In this section we
concentrate on the
normal state only. Thus, Eqs.~\eqref{eq:23s} %and \eqref{eq:24}
apply with
$\alpha^2F(\omega)$ replaced by Eq.~\eqref{eq:29}. Results for a case
with $\omega_{SF}=20\,$meV, $I^2=0.82$, and $D/2 = 400\,$meV are shown in
Fig.~\ref{fig:10} which has three frames. In the top frame we show
$-\Sigma_2(T,\omega)$ for four temperatures as labeled, in the center
frame $\Sigma_1(T,\omega)$,
 and in the bottom frame $h(T,-\omega)$
vs $\omega$. It is clear that both real and imaginary part of
$\Sigma(T,\omega)$ have important $T$ dependencies which get
reflected in $h(T,\omega)$ and consequently in the OS given by
Eq.~\eqref{eq:17} and more approximately by Eq.~\eqref{eq:19} with
\begin{figure}[tp]
%  \centering
  \includegraphics[width=9cm]{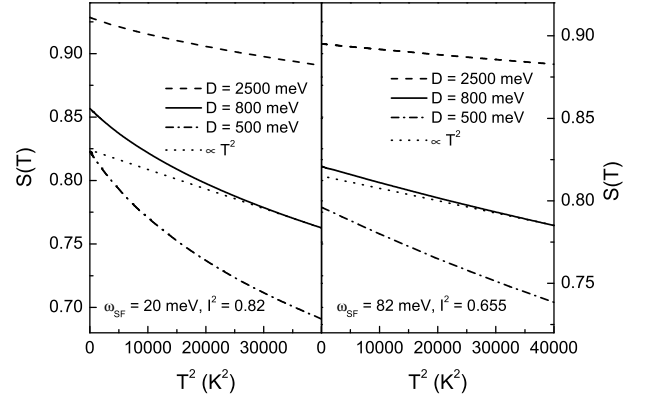}
  \caption{The reduced optical sum $S(T)$ according to
Eq.~\protect{\eqref{eq:29a}}
vs temperature $(T)$ squared. The left hand frame is for
an MMP model \protect{\eqref{eq:29}} with $\omega_{SF} = 20\,$meV and
$I^2=0.82$ while the right hand frame is for $\omega_{SF} = 82\,$meV
and $I^2 = 0.655$.
}
  \label{fig:11}
\end{figure}
the last term dropped. In this instance it is simply the area under
$h(T,\omega)$ for negative $\omega$ that determines the OS while in
Eq.~\eqref{eq:17} it is its overlap with the Fermi distribution and this
has an additional temperature dependence due to $f(T,\omega)$ but it
is small. Results for
\begin{equation}
   \label{eq:29a}
   S(T) = 2\bar{W}(T)\frac{m_b}{\hbar^2}
\end{equation}
are shown in Fig.~\ref{fig:11} which has two frames. In the left hand
frame we show results for $S(T)$ vs $T^2$ up to $40 000\,$K$^2$ for our
MMP model with $\omega_{SF} = 20\,$meV and $I^2 = 0.82$ while in the
right hand frame $\omega_{SF}$ is increased to $82\,$meV with
$I^2=0.655$. In both cases the full band width $D$ ranges over 2500
(dashed line), 800 (solid line), and $500\,$meV (dash-dotted line).
It is clear that decreasing $D$ reduces the value of $S(T=0)$ and
gives a stronger $T$ dependence which, nevertheless, remains near
a $T^2$ behavior for the right hand frame but not in the left hand
frame. These results are in qualitative agreement with those shown
in the previous section, Figs.~\ref{fig:1} and \ref{fig:3} obtained
in a tight binding model with a {\bf k}-point sampling method of
the Brillouin zone and the full momentum dependent Eliashberg
equations \eqref{eq:6}. Also shown in both frames as a dotted curve for the
case $D=800\,$meV are straight lines which allow us to judge better
the deviations from a $T^2$ law at small $T$ in the curves of the left
hand frame. For the right hand frame there is little deviation.
Note that both these behaviors are completely consistent with
numerical results of Fig.~\ref{fig:3} and show that the upward
deviation at small $T$ of the solid curve as compared with the
dotted $(T^2)$ is a robust result not dependent on the model
used, provided that the spin fluctuation frequency is small.
Note also that Eq.~\eqref{eq:17} includes $T$ variations in $h(T,\omega)$
as well as the thermal factor $f(T,\omega)$, but this latter factor
can be neglected. The differences are very small
confirming our previous claim that the temperature dependence
in $S(T)$ is mainly due to variations in the interaction term
at least for these cases. Finally, we stress again that the $T^2$
law found in the DMFT calculations for the Hubbard model and
often confirmed in experiments also arises in the NAFFL model with
an MMP interaction provided the spin fluctuation energy is fairly large
and $D$ not too small. In principle, however, there is no fundamental
reason for an exact $T^2$ law. The actual $T$ dependence of $S(T)$
comes from the $T$ dependence of the area under the curves for
$h(T,-\omega)$ (bottom frame of Fig.~\ref{fig:10}). We see that
these curves change most at small $\omega$ as $T$ is increased but
there are also important changes at large $\omega$ and we are sampling
an average of these changes. Nevertheless, the $\omega$
dependence of $\alpha^2F(\omega)$ or $I^2\chi(\omega)$ at low energies
$\omega$ will determine most strongly the temperature dependence of
$h(T,-\omega)$ at those energies and, thus, plays an important role
in the temperature dependence of $S(T)$. However, even at higher
energies there is still significant $T$ dependence in $h(T,-\omega)$.

\subsection{The non $T^2$ temperature dependence of the optical sum}
\label{ssec:3e}

Benfatto {\it et al.}\cite{ref52} have reconsidered the effect of
\begin{figure}[tp]
%  \centering
%  \includegraphics[width=9cm]{fig3bw.eps}
  \includegraphics[width=8cm]{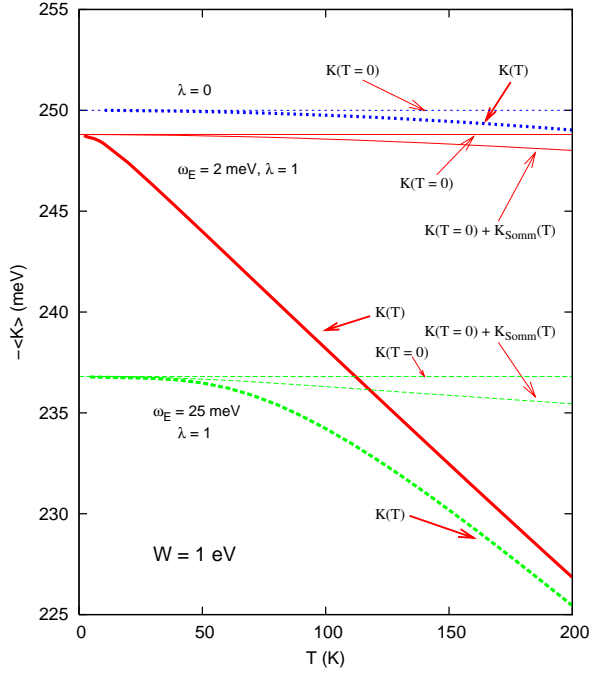}
  \caption{(Color online)
The negative of the KE, $-\langle K\rangle$, vs temperature
showing the breakdown for the various contributions. The solid
curves are for a small boson frequency $\omega_E=2\,$meV with
$\lambda=1$. The two thin lines show how the zero temperature value
is modified by the Sommerfeld term by a very small amount. A similar
size contribution (actually a bit larger) is illustrated for the
noninteracting case by dotted curves, where only the
Sommerfeld term is responsible for the temperature variation. Finally,
for higher boson frequencies, the dashed curves illustrate
the amount of temperature variation due to the Sommerfeld term
compared with the rest. The band width $W \equiv D = 1\,$eV.
 Adapted from Ref.~\protect{\onlinecite{ref52}}.
}
  \label{fig:12}
\end{figure}
interaction on the temperature dependence of $S(T)$ with an aim at
providing simple analytic expressions that would help trace more
directly the dependence on the interactions.
In Fig.~\ref{fig:12} we reproduce some of their results for the
temperature dependence of the negative of the KE, $-\langle K\rangle \equiv
K(T)$, in meV. The two dotted lines at the top of the graph represent
the noninteracting case $(\lambda=0)$ and are for reference. As expected,
there is little difference between $K(T)$ and its zero temperature
value $K(T=0)$. The three solid curves are for a system with coupling
to a boson of frequency $\omega_E=2\,$meV and $\lambda=1$. We see that
$K(T=0)$ and $K(T=0)+K_{\rm Somm}(T)$, the Sommerfeld contribution, are
not significantly different. Thus, $K_{\rm Somm}(T)$ is not the
important contribution to the temperature dependence of the complete KE
[heavy solid line, denoted $K(T)$] which develops a linear temperature
dependence for $T\sim20\,$K. The same holds for the case
$\omega_E=25\,$meV (three dashed curves) but the variation in $T$ of
$K(T)$ in this case is less pronounced and closer to a $T^2$ law as
was found for the MMP spectrum. The band width in all three cases was
$W \equiv D = 1\,$eV. Benfatto {\it et al.}\cite{ref52} trace these
$T$ dependences in detail using analytic as well as numerical techniques.
They concluded that both the real and imaginary parts of $\Sigma$
contribute. Karakozov and Maksimov\cite{ref53} also find deviations
from a $T^2$ law and provide an approximate analytic formula for
these deviations in a particular case.
For different models of $\alpha^2F(\omega)$ other
$T$ dependencies are possible. This means that in principle one
can learn about some features of this underlying $\alpha^2F(\omega)$
from a study of the temperature dependence of the OS, but the
correspondence is not necessarily simple or unique.
On the other hand, a complete study of the
temperature and frequency dependence of the optical self energy\cite{ref64}
as is now done routinely, provides much more detailed information
than does the OS.
Recall that $h(T,\omega)$ is dependent on an average over the
real and imaginary part of $\Sigma$, the quasiparticle self energy.
This information is accessible directly from angular resolved
photoemission spectroscopy (ARPES) at each point in the Brillouin
zone separately.

\begin{figure}[tp]
\vspace*{-16mm}
%  \centering
%  \includegraphics[width=10cm]{srFig13.eps}
  \includegraphics[width=9cm]{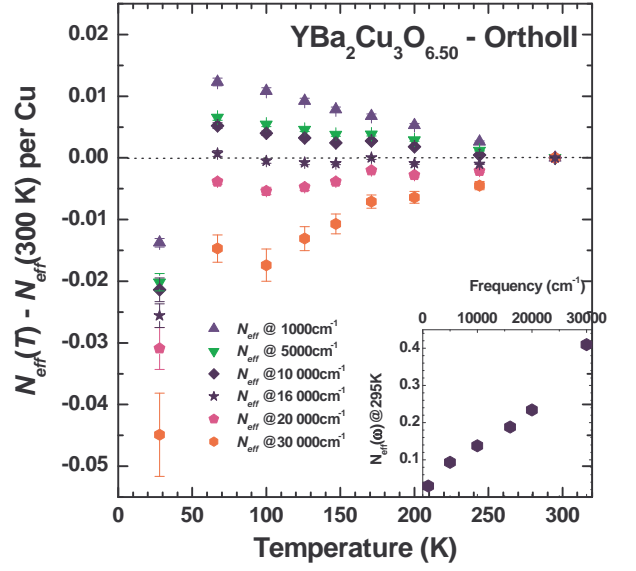}
  \vspace*{-2cm}
  \caption{(Color online)
The partial spectral weight integrated up to various
frequencies as a function of temperature. Below $16\,000\,$cm$^{-1}$
there is an increase in spectral weight as the temperature is lowered
signaling a line narrowing on this frequency scale. Below $T_c$ there
is strong loss of spectral weight to the superconducting condensate.
There is no evidence of any precursors to superconductivity at
$67\,$K. In the inset we show $N_{eff}(\omega)$ at $295\,$K. Adapted
from Ref.~\protect{\onlinecite{ref64}}.
}
  \label{fig:13}
%\vspace*{-1cm}
\end{figure}
While many experiments give a $T^2$ dependence in the normal state
within the precision of measurement there is some evidence for
other dependences in the cuprates. In Fig.~\ref{fig:13} we reproduce
data from the paper of Hwang {\it et al.}\cite{ref64} for the partial sum
to $\omega_c$ in units of carriers per Cu atom denoted by
$N_{eff}(T)$. What is shown is the difference $N_{eff}(T)-N_{eff}(300\,
{\rm K})$ as a function of temperature $T$ to $300\,$K. The data is for
underdoped
YBa$_2$Cu$_3$O$_{6.50}$ (YBCO$_{6.50}$) Ortho II material of high
quality and purity. There are 5 values of $\omega_c$, namely 1000, 5000,
$10\,000$, $16\,000$, and $20\,000\,$cm$^{-1}$. It is clear that the
temperature dependence of $N_{eff}(T)$ is not necessarily quadratic
in $T$ for the partial sums and also for the case most relevant for
our discussions with $\omega_c = 16\,000\,$cm$^{-1}$. These samples
are, however, underdoped and a pseudogap is involved which could
have some effect on the temperature dependence of the OS as we will
discuss later within a phase fluctuation model for the pseudogap and
alternatively a $D$-density wave model.

\section{The Optical Sum in the Superconducting State}
\label{sec:4}
\subsection{Superconducting optical sum including of inelastic
scattering collapse}
\label{ssec:4a}

\begin{figure}[tp]
\vspace*{-4mm}
  \includegraphics[width=8.5cm]{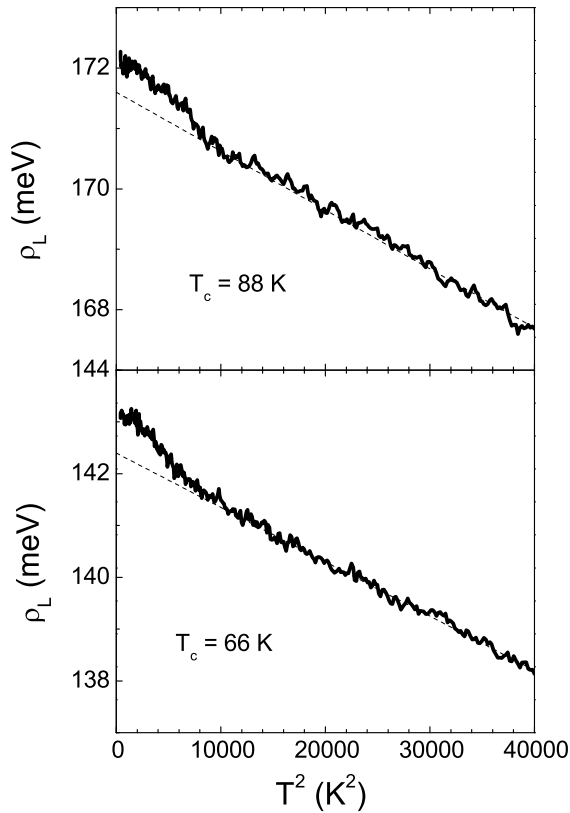}
  \caption{Experimental values of the $ab$-plane spectral function 
$\rho_L$ defined in Eq.~\protect{\eqref{eq:30}}.
Adapted from Ref.~\protect{\onlinecite{ref34}}.
}
  \label{fig:14}
%\vspace*{5cm}
\end{figure}
Next we return to a more detailed look at the superconducting state.
In Fig.~\ref{fig:14} we reproduce the original results of
Molegraaf {\it et al.}\cite{ref48} for two samples of BSCCO
as they were presented by van der Marel {\it et al.}\cite{ref34}
The top frame is for optimally doped and the bottom frame for
an underdoped sample. What is shown is
\begin{equation}
  \label{eq:30}
  \rho_L \equiv C\frac{\hbar^2}{\pi e^2}\int\limits_{-\Omega}^\Omega\!
d\omega\,\Re{\rm e}[\sigma(\omega)]
\end{equation}
in meV with $C$ a constant to be specified shortly.
It is striking that in these two samples the OS appears to
increase faster in the superconducting state than in the underlying
normal state extrapolated to low temperatures (dotted line). This
was interpreted by Molegraaf {\it et al.}\cite{ref48} as indicating
a decrease in KE in the superconducting state as compared with the
underlying normal state which also shows a decrease in KE as $T\to 0$
but by less than in the superconducting state. This behavior
clearly goes beyond ordinary BCS theory and also Eliashberg theory as
formulated for phonons.

There remains some controversy about the interpretation of such data
with some authors finding the opposite effect.\cite{ref65} We will not go into
these details here but refer the reader to some relevant
literature\cite{ref65,ref66,ref67} and note that Santander-Syro%
\cite{ref47} confirm the basic results of Molegraaf {\it et al.}\cite{ref48}
See also the recent work of Carbone {\it et al.}\cite{ref33a}

We have already seen in Fig.~\ref{fig:9} [inset in frame (a)] for
the normal state, that undressing effects associated with a hardening of
the phonon spectra at a particular onset temperature can give an OS curve
that behaves very much like those seen in Fig.~\ref{fig:14}.
While in principle
phonon frequencies can shift as a result of the onset of superconductivity
these effects are small and usually negligible. For an electronic mechanism,
however, we have already discussed in the introduction the idea of the
collapse of the inelastic scattering as superconductivity sets in which
leads directly to a peak in the real part of the microwave conductivity
at some intermediate temperature below $T_c$. Within a spin fluctuation
mechanism this translates into a hardening of the spin fluctuation spectrum
and the modification of $I^2\chi(\omega)$ of Eq.~\eqref{eq:29} as the
$d$-wave superconducting gap opens. An analysis of the optical scattering
rates in YBCO$_{6.95}$ by Carbotte {\it et al.}\cite{ref68} showed that
\begin{figure}[tp]
\vspace*{7mm}
\vspace*{-2mm}
  \includegraphics[width=8.5cm]{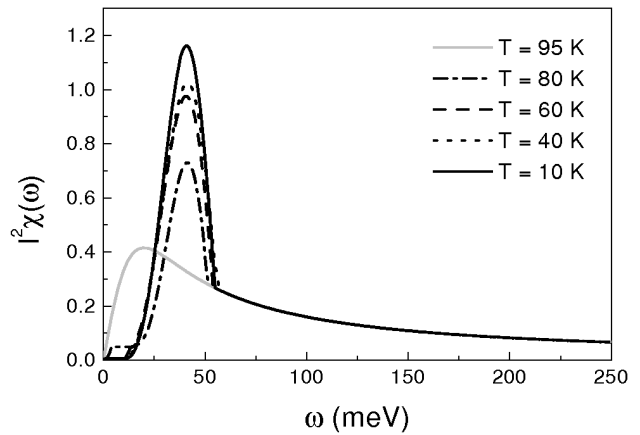}
  \caption{The charge carrier-spin spectral density $I^2\chi(\omega)$
  determined from optical scattering data of YBCO$_{6.95}$
at various temperatures.
  Solid gray curve $T=90\,$K, dash-dotted $T=80\,$K, dashed $T=60\,$K,
dotted $T=40\,$K, and black solid $T=10\,$K. Note the growth in strength
of the $41\,$meV optical resonance as the temperature is lowered.
Adapted from Ref.~\protect{\onlinecite{ref70}}.
}
  \label{fig:15}
\end{figure}
there is a reduction in $I^2\chi(\omega)$ at small $\omega$ as the
temperature is reduced. Results obtained by Schachinger
{\it et al.}\cite{ref69,ref70} for the temperature evolution of
$I^2\chi(\omega)$
in YBCO$_{6.95}$ are reproduced in Fig.~\ref{fig:15}. We see a reduction in
spectral weight at low $\omega$ as well as the growth of an optical
resonance at $41\,$meV, the energy of the spin one resonance seen in
neutron scattering.\cite{ref71} When these spectra are used in a
\begin{figure}[tp]
%  \centering
\vspace*{8mm}
  \includegraphics[width=8cm]{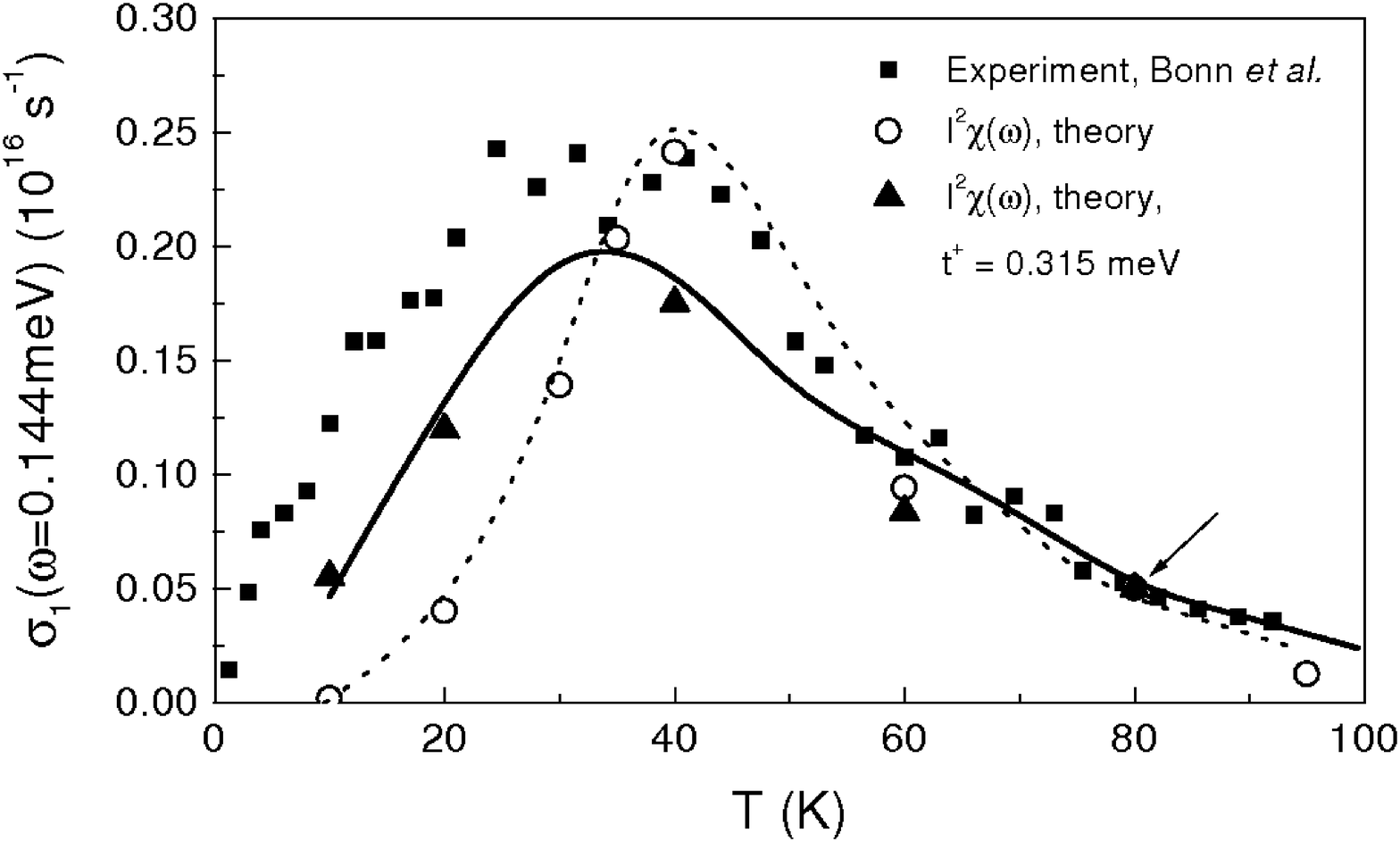}
  \caption{
Temperature dependence of the conductivity $\sigma_1(\omega)$ at
microwave frequency $\omega=0.144\,$meV in YBCO$_{6.95}$.
The open circles are results based on
our model spectral density (see Fig.~\protect{\ref{fig:15}}) obtained
from inversion of optical conductivity data and the dashed line is
based on an
MMP model with low frequency cutoff. The solid triangles include impurities
with the solid line from Ref.~\protect{\onlinecite{ref74}}. The solid
squares represent experimental data by Bonn {\it et al.}\protect{%
\cite{ref75}} Adapted from Ref.~\protect{\onlinecite{ref69}}.
}
  \label{fig:16}
\end{figure}
generalization of the ordinary Eliashberg equations that
includes the possibility of $d$-wave symmetry with gap $\Delta(\phi) =
\Delta_0\cos(2\phi)$ where $\phi$ is an angle on the Fermi surface
taken to be cylindrical, Schachinger and Carbotte,\cite{ref70,ref72}
find excellent agreement for the microwave data of Hosseini {\it et al.}%
\cite{ref73} taken at five different microwave frequencies on high quality
samples of YBCO$_{6.99}$. In Fig.~\ref{fig:16}, reproduced from
Ref.~\onlinecite{ref69}, we show older results obtained by Bonn {\it et
al.}\cite{ref75} fit by Schachinger {\it et al.}\cite{ref74} by simply
applying a low frequency cutoff to an MMP form.\cite{ref74,ref76} The data
are given as the solid squares while the theoretical calculations without
impurities (open circles and dashed line) and with impurities
(solid triangles and solid line) are shown for comparison. Here
$t^+$ gives the impurity scattering rate in Born approximation.
Open circles and solid triangles are based on the spectra of Fig.~\ref{fig:15}
which include the $41\,$meV spin resonance. The dashed and solid lines
are from earlier calculations based on an MMP spectrum with application
of a low frequency cutoff without resonance.\cite{ref74,ref76} We see
that such a procedure gives results that are very close to those based
on the more exact results of Fig.~\ref{fig:15}.

Calculations in the model of Eqs.~\eqref{eq:6} of the OS and KE
of optimally doped BSCCO have
been carried out by Schachinger and Carbotte\cite{ref40} who simulated
\begin{figure}[tp]
%  \centering
  \includegraphics[width=9cm]{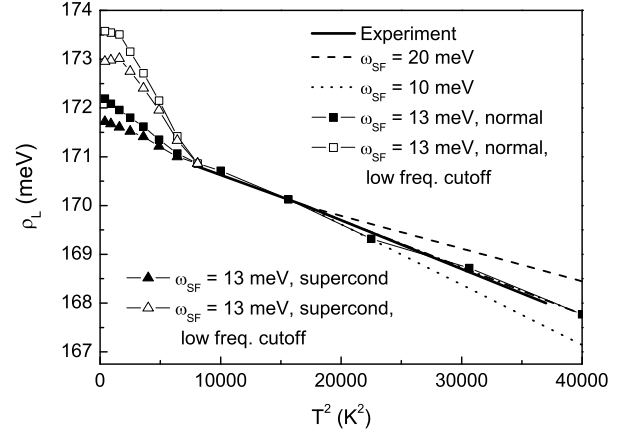}
  \caption{The optical sum of optimally doped BSCCO 
as a function of the square of the temperature
for the band structure Model A of Table~\protect{\ref{tab:1}} with
different values of $\omega_{SF}$. Also in one case a low frequency
cutoff is applied to the spin susceptibility. Note the significance
of the $T^2$-variation on the value of $\omega_{SF}$. The thick solid
line represents experimental normal state data of Molegraaf {\it et al.}%
\protect{\cite{ref48}} also shown in the top frame of
Fig.~\protect{\ref{fig:14}}.
Adapted from Ref.~\protect{\onlinecite{ref40}}.
}
  \label{fig:17}
\end{figure}
the expected hardening of the spin fluctuation spectra in the
superconducting state by applying a low frequency cutoff to the
interaction of Eq.~\eqref{eq:7}. This cutoff varies with temperature.
It is zero at $T_c$ and has a maximum amplitude of $23\,$meV at $T=0$ with
intensity changed according to a mean field BCS $d$-wave order parameter
temperature dependence. Their results are shown in Fig.~\ref{fig:17}.
The quantity $\rho_L$ according to Eq.~\eqref{eq:30} on the vertical
axis is in meV and the horizontal scale is $T^2$ in K$^2$. The heavy
solid line are normal state experimental data of Molegraaf {\it et al.},%
\cite{ref48} also shown in the top frame of Fig.~\ref{fig:14} of this
review. The theoretical results for the OS are all based on Model A
of Table~\ref{tab:1} but were scaled upward by a factor
[$C$ of Eq.~\eqref{eq:30}] of approximately
two to fit the data. The spin fluctuation frequency $\omega_{SF}$
was also adjusted to improve the fit; $\omega_{SF}=13\,$meV is best.
To reduce the discrepancy in absolute value of the OS the value of
the nearest neighbor hopping parameter $t$ in our band structure model
would need to be increased. This would, however, also decrease the
sensitivity of the OS to temperature variations and so $\omega_{SF}$
would have to be adjusted downwards as well. Markiewicz {\it et al.}%
\cite{ref77} have suggested significantly larger values of nearest
neighbor hopping $t$ for
the bare band structure of BSCCO  than used to fit ARPES data. As is
well known, there can be a factor of two or more. We have
used the dispersion relation of Table I of Ref.~\onlinecite{ref77}
(specifically we used $t=360\,$meV, $t'=-100\,$meV, and
$\langle n\rangle = 0.4$. We also included further neighbors, namely
$t''=35\,$meV and $t'''=20\,$meV)
and found a value of $W(T=0)$ of $230\,$meV well above the
experimental value with $\omega_{SF}=8\,$meV giving the proper
temperature dependence. It is clear that a value of $t$ intermediate between
that of Markiewicz {\it et al.}\cite{ref77} and the one in Table~\ref{tab:1}
for Model A would be needed to get agreement with the OS data without any
adjustment and a $\omega_{SF}$ value between 8 and $13\,$meV. Schachinger
and Carbotte\cite{ref40} did not attempt such a fit
as their aim was not
to fit any particular case but rather see how a low frequency cutoff
applied to a spin fluctuation model changes the OS. The solid
squares in Fig.~\ref{fig:17} show the normal state results while the
solid triangles are in the superconducting state. These are obtained
without cutoff and so the KE has increased with respect to its normal
state. The open squares and triangles show results for the same two
cases but now the low frequency cutoff is applied to simulate the
hardening of the spectrum as $T$ is reduced below $T_c$. We see a
large decrease in KE as compared to the case without cutoff and the
superconducting state would then show a decrease in KE as compared to
the normal state without cutoff in agreement with the data of
Fig.~\ref{fig:14}. It is the modification of the underlying interactions
that have lead to this effect. It is clear that a modest hardening of
the spectra consistent with the changes seen in the electron-boson
spectral functions of Fig.~\ref{fig:15} would be sufficient to explain
the data in Fig.~\ref{fig:14}. Of course, in a complete theory as yet
not attempted, it would be necessary to find an interaction which
can give not only the correct OS but also the microwave peak
and all the
other properties of the superconducting state. Finally, we note that
in Fig.~\ref{fig:17} the OS shows a non $T^2$ behavior at low
temperatures even without the low frequency cutoff which has the effect
of making it more pronounced.
If we did not have the normal state data below $T_c$ and simply
extrapolated the normal state data above $T_c$ with a $T^2$ law 
to $T=0$ we would have to conclude that the OS is higher in
the superconducting state
than in the extrapolated ``normal state'' yet in this case the mechanism
for pairing is not exotic in any way, i.e.: there is no low frequency cut
off. (See also Fig.~\ref{fig:3}.) With the introduction of a low frequency
cutoff this effect becomes even more pronounced and
for the case shown the rise in the superconducting state is larger
than seen in the experiments of Fig.~\ref{fig:14}. We caution the reader,
however, that while this rise is seen by several experimental groups and is
not controversial, its actual size is.

\subsection{The Temperature Dependent Scattering Time Model}
\label{ssec:4c}

Recently Marsiglio\cite{ref90} has
given calculations that provide support for the results of Fig.~\ref{fig:17}
using a related but much simpler model. He notes that in an infinite band for
an elastic scattering rate $\Gamma$ the probability of occupation of the
state $\vert{\bf k}\rangle$ at finite temperature is given by
\begin{equation}
  \label{eq:32}
  n_N(\epsilon_{\bf k}) = \frac{1}{2}-\frac{1}{\pi}\,\Im{\rm m}\psi\left(
  \frac{1}{2}+\frac{\Gamma}{4\pi T}+i\frac{\epsilon_{\bf k}}{2\pi T}
  \right)
\end{equation}
for the normal state and, to a good approximation, by
\begin{equation}
  \label{eq:33}
  n_S(\epsilon_{\bf k}) = \frac{1}{2}-\frac{\epsilon_{\bf k}}{E_{\bf k}}
  \frac{1}{\pi}\Im{\rm m}\,\psi\left(
  \frac{1}{2}+\frac{\Gamma}{4\pi T}+i\frac{E_{\bf k}}{2\pi T}
  \right)
\end{equation}
for the superconducting state with $E_{\bf k} = \sqrt{\epsilon_{\bf k}^2+
\Delta_{\bf k}^2}$,
with $\Delta_{\bf k}$ the superconducting gap and
$\psi(z)$ is the digamma function.  Note that at zero temperature
\begin{equation}
  \label{eq:34}
  n_S(\epsilon_{\bf k}) = \frac{1}{2}\left[1-\frac{\epsilon_{\bf k}}
  {E_{\bf k}}\frac{2}{\pi}\tan^{-1}\left(\frac{2E_{\bf k}}{\Gamma}\right)
  \right],
\end{equation}
which, in the limit $\Gamma\to 0$, reduces to the well known expression
$(1-\epsilon_{\bf k}/E_{\bf k})/2$, Eq.~\eqref{eq:11a}
introduced in Sec.~\ref{ssec:2d}.
It is clear from Eq.~\eqref{eq:34} that both, the appearance of a
gap in $E_{\bf k}$ and the elastic scattering $\Gamma$ smear
$n_S(\epsilon_{\bf k})$.
\begin{figure}[tp]
%  \centering
%  \includegraphics[width=11cm]{srFig21.eps}
  \includegraphics[width=9cm]{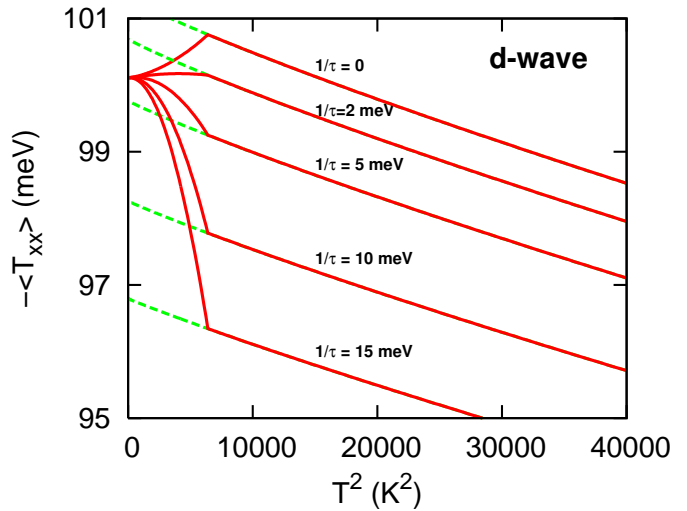}
  \caption{(Color online)
Minus the kinetic energy, $-\langle T_{xx}\rangle$,
vs temperature squared for various
degrees of elastic scattering. Below $T_c$ we change the elastic
scattering rate smoothly to zero [as $\Gamma_0 (T/T_c)^4$] as the
temperature is lowered.
Adapted from Ref.~\protect{\onlinecite{ref90}}
}
  \label{fig:21}
\end{figure}
From these expressions for $n(\epsilon_{\bf k})$ it is easy
to work out the average KE which Marsiglio denotes by $-\langle T_{xx}
\rangle$. He uses a tight binding band and a constant
$\Gamma_0 = 1/\tau = 0$, 2, 5, 10, and $15\,$meV from top to bottom
in Fig.~\ref{fig:21}. For the superconducting state $\Gamma_0$ is
changed to $\Gamma(T)$ with $\Gamma(T) = \Gamma_0(T/T_c)^4$ which is
observed in the work of Hosseini {\it et al.}\cite{ref73} This
introduces phenomenologically the idea of the collapse of the
scattering rate. We see in Fig.~\ref{fig:21} that for $1/\tau = 5\,$meV
and larger the KE decreases below its normal state extrapolation as the
system becomes superconducting. The change is of the order of a few
percent for the largest $1/\tau$ considered and is large enough to
explain the measured KE changes in the underdoped cuprates.

\subsection{The model of Norman and P\'epin}
\label{ssec:4b}

The mechanism described above which leads to an OS increase as
superconductivity sets in has some common elements with the ideas
of Norman and P\'epin\cite{ref78,ref79} although there are also
important differences. These authors use a less fundamental, more
phenomenological approach based on ARPES\cite{ref80} and optical
conductivity data from which they construct directly a model for
the self energy $\Sigma(\omega)$. For the normal state they begin
with a constant, frequency independent scattering rate
$\Gamma_0 = \Im{\rm m}\Sigma(\omega)$ (taken to be of order
$100\,$meV) to simulate the very broad (incoherent) line shapes seen
in ARPES at the anti nodal point $(0,\pi)$. In the superconducting
state a sharp coherence peak appears which signals increased coherence.
To model this fact a low frequency cutoff $\omega_0$ is applied
to the imaginary part of $\Sigma(\omega)$ making it zero for
$\omega\le\omega_0$ and $\Gamma_0$ above. The value of $\omega_0$
is chosen as the energy of the spectral dip feature seen in ARPES
in the superconducting state. Here, as in the model described
in Sec.~\ref{ssec:4c},
the underlying normal state on top of which superconductivity
develops effectively has reduced scattering, i.e.: is more coherent
which is a critical additional feature not contained in ordinary
BCS and this leads to a reduction in KE and, consequently, in an increase
of the OS. To arrive at their final estimates for the KE increase
Norman and P\'epin include further complications in their model
such as the anisotropy observed in scattering rates\cite{ref81,ref82,ref83,%
ref84} ($\Gamma_{\bf k}$ instead of $\Gamma_0$)
\begin{figure}[tp]
%  \centering
%\vspace*{5mm}
  \includegraphics[width=7cm]{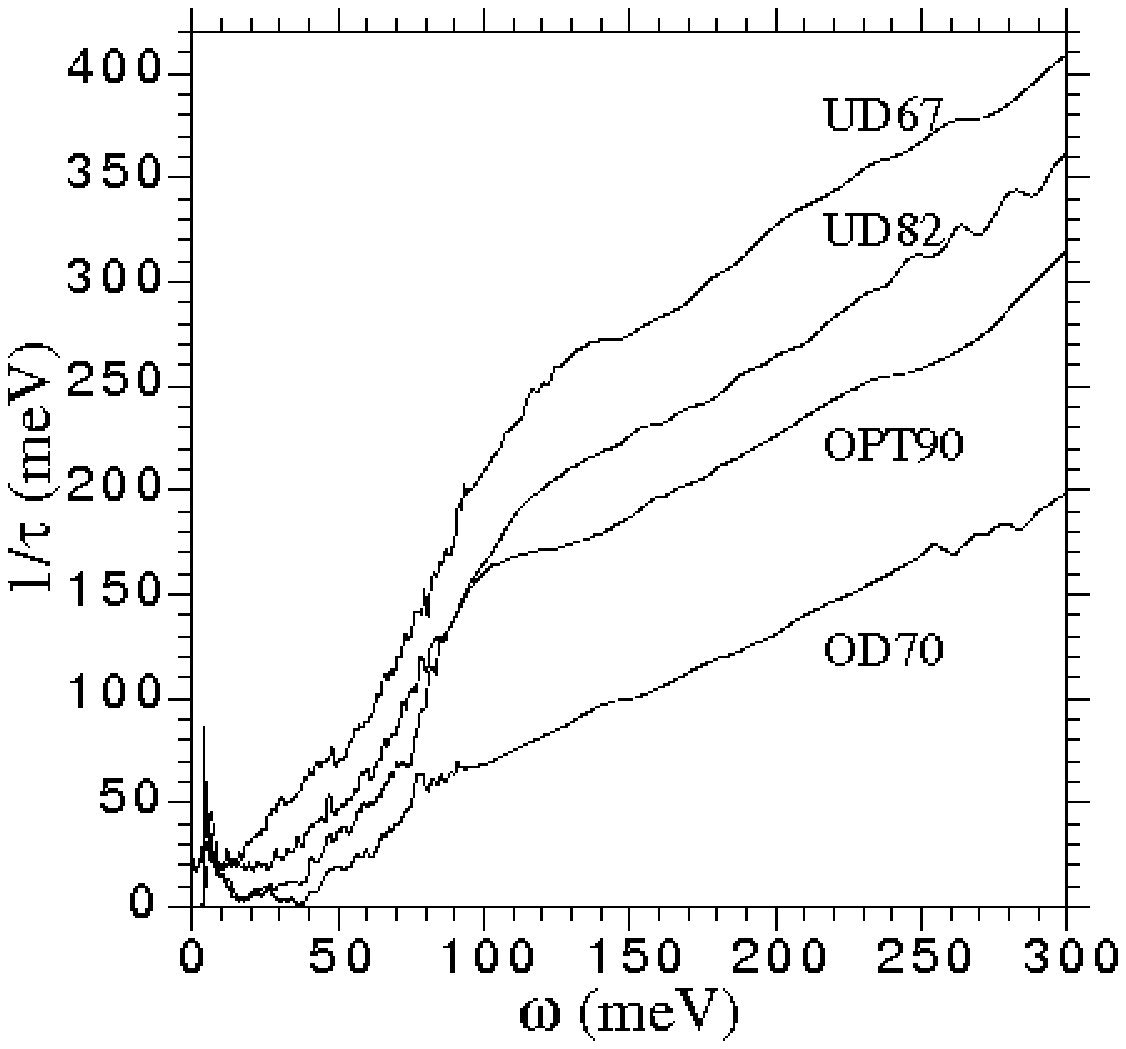}
  \includegraphics[width=7cm]{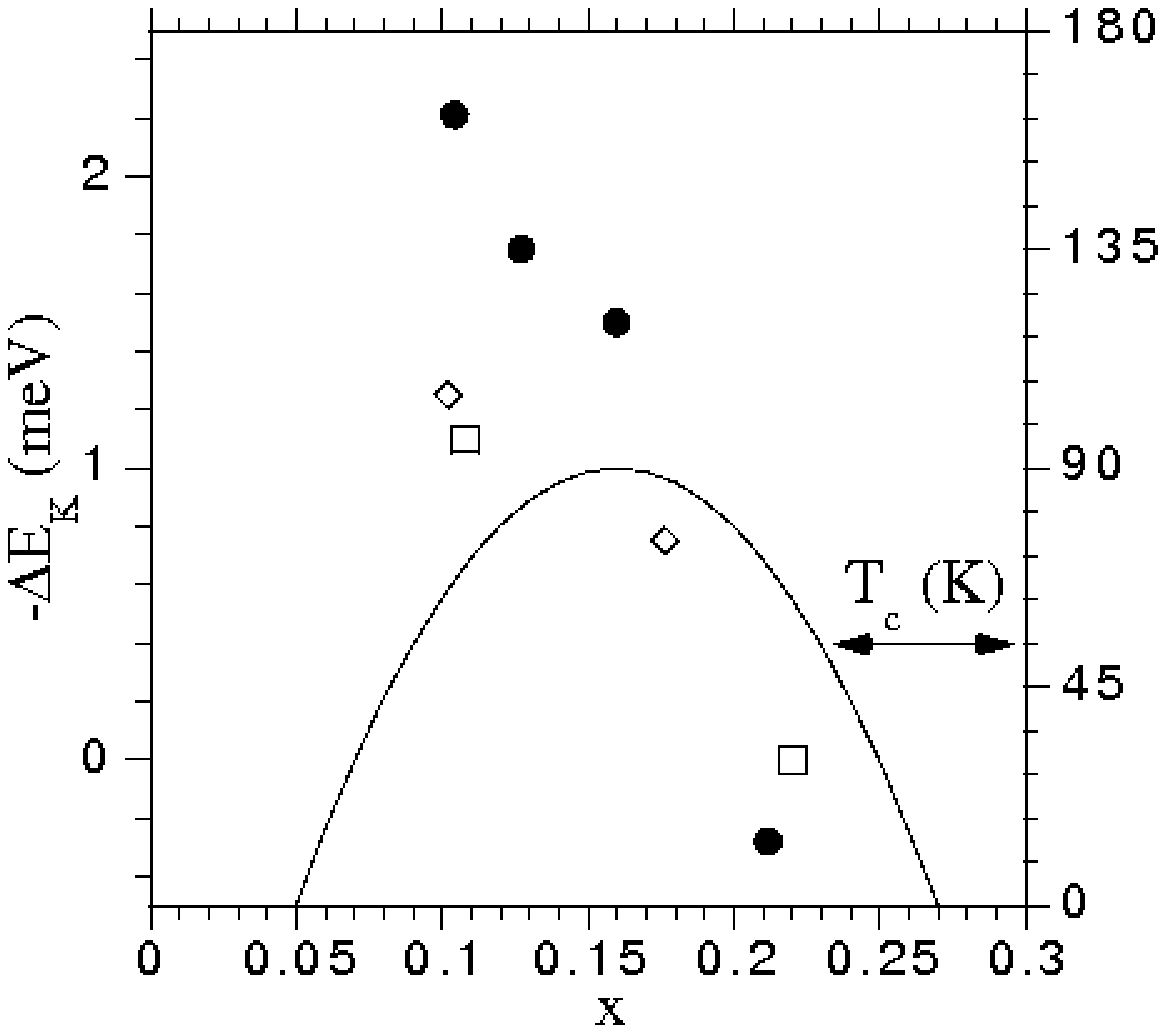}
  \caption{Top frame:
The optical scattering rate $1/\tau(\omega)$ vs $\omega$
for various BSCCO samples from
Ref.~\protect{\onlinecite{ref87}} (OD overdoped, OPT optimal doped,
UD underdoped). Bottom frame:
Calculated sum rule violation $(-\Delta E_K)$ vs
doping $x$ (solid circles).
The curve is $T_c$. Also shown in this frame are the
experimental results (open squares from Ref.~\protect{\onlinecite{ref46}},
open diamonds from Ref.~\protect{\onlinecite{ref48}}). The theoretical
doping trend in (b) is due to the increasing offset in $1/\tau$ seen
in (a). Adapted from Ref.~\protect{\onlinecite{ref78}}.
}
  \label{fig:18}
\end{figure}
going from anti nodal to nodal direction as well as an
$\omega$ dependence (proportional to a momentum independent parameter
$\alpha$) in the self energy modeled on the Marginal
Fermi Liquid model\cite{ref85,ref86} with parameters determined
from optical
data on scattering rates. We reproduce in Fig.~\ref{fig:18} their
final estimates for the KE change $-\Delta E_K$ between superconducting
and normal state. The left hand frame shows the optical scattering rate
denoted $1/\tau(\omega)$ for four
BSCCO samples from Puchkov {\it et al.}\cite{ref87} which they use
to fit parameters. The right hand frame presents the results for the change in
KE $(-\Delta E_K)$ associated with the formation of the superconducting
state in
meV as a function of doping $x$. The solid circles give the theoretical
results and the open squares and open diamonds experimental data from
Santander-Syro {\it et al.}\cite{ref46} and Molegraaf {\it et al.},\cite{ref48}
respectively. The theoretical estimates are deemed reasonable
but not accurate. On the overdoped side the OS behaves in a conventional
fashion while for the underdoped side it is anomalous representing a
lowering of KE as compared with what is expected in conventional BCS.
Note that the lowest open square in this graph shown as being zero for
the overdoped sample represents an early not very accurate estimate. In
Ref.~\onlinecite{ref46} it is actually negative rather than zero
bringing it closer in agreement with the lowest solid circle (theory).
This point was further discussed by Deutscher {\it et al.}\cite{ref87a}
%taken from
%Ref.~\onlinecite{ref48} for an overdoped sample falls at negative values
%of $-\Delta E_K$ rather than at zero\cite{ref87a,ref88} as shown in
%Fig.~\ref{fig:18} bringing it in closer agreement with the lowest solid
%circle (theory).

\subsection{Additional data on kinetic energy changes}
\label{ssec:4e}

\begin{figure}[tp]
%  \centering
  \includegraphics[width=8cm]{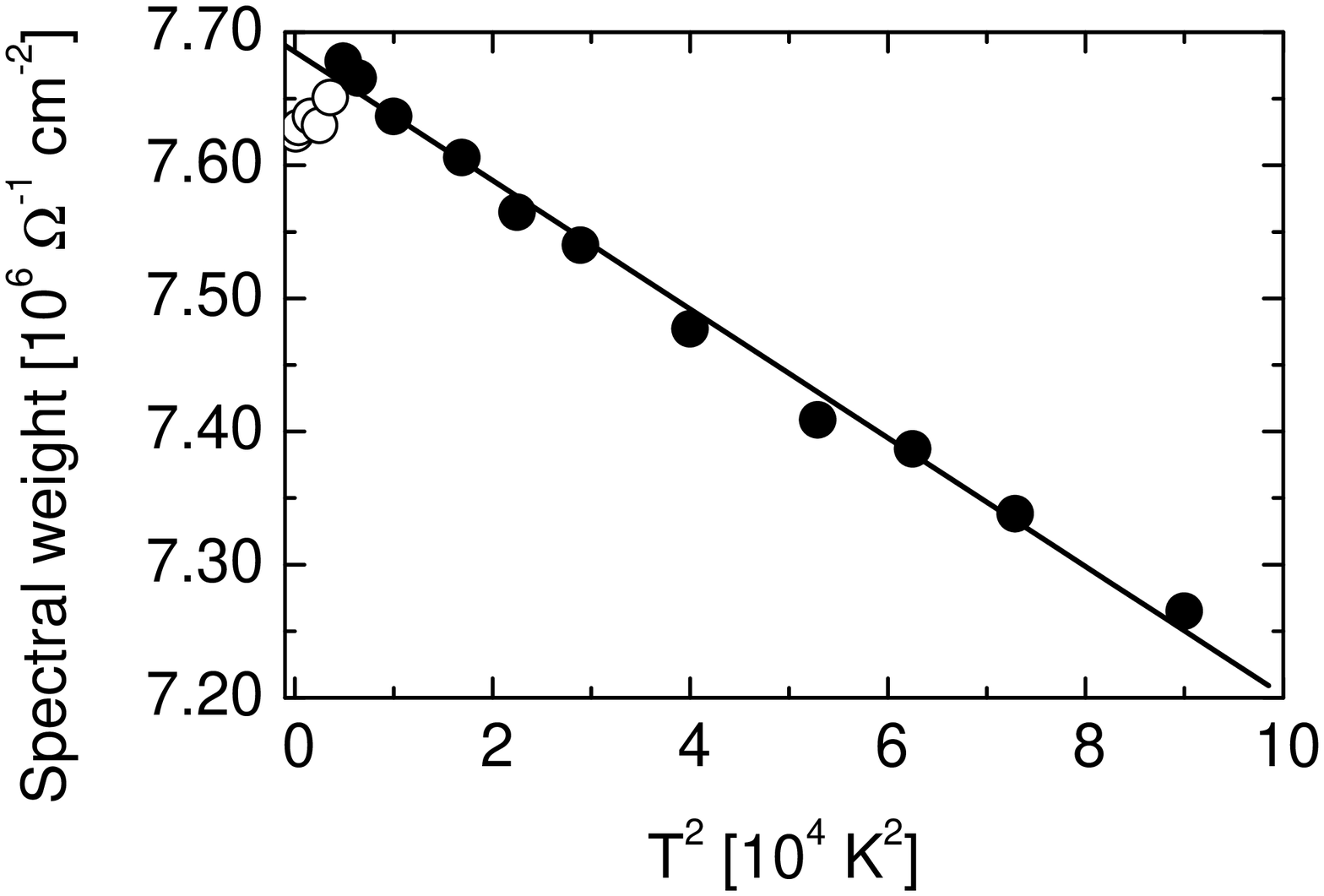}
  \caption{Spectral weight of the overdoped Bi2212 sample, integrated
up to $1\,$eV, plotted vs $T^2$, from Ref.~\protect{\onlinecite{ref46}}.
Closed symbols: spectral weight in the normal state, open symbols:
spectral weight in the superconducting state, including the weight
of the superfluid. The errors in the {\it relative} variations of the
spectral weight are of the size of the symbols. Adapted from
Ref.~\protect{\onlinecite{ref87a}}.
}
  \label{fig:19}
\end{figure}
We reproduce in Fig.~\ref{fig:19} the data of Santander-Syro {\it et al.}%
\cite{ref46} for their overdoped sample as presented by Deutscher
{\it et al.}\cite{ref87a} What is shown is the optical spectral weight
in units $10^6\,\Omega^{-1}\,{\rm cm}^{-2}$ as a function of $T^2$.
The solid circles are in the normal state and the open circles in the
superconducting state below $T_c = 63\,$K. A clear $T^2$ law is noted
above $T_c$ and a reduction in spectral weight below, as expected in
ordinary BCS theory.  For the optimally doped sample
\begin{figure}[tp]
%  \centering
  \includegraphics[width=8cm]{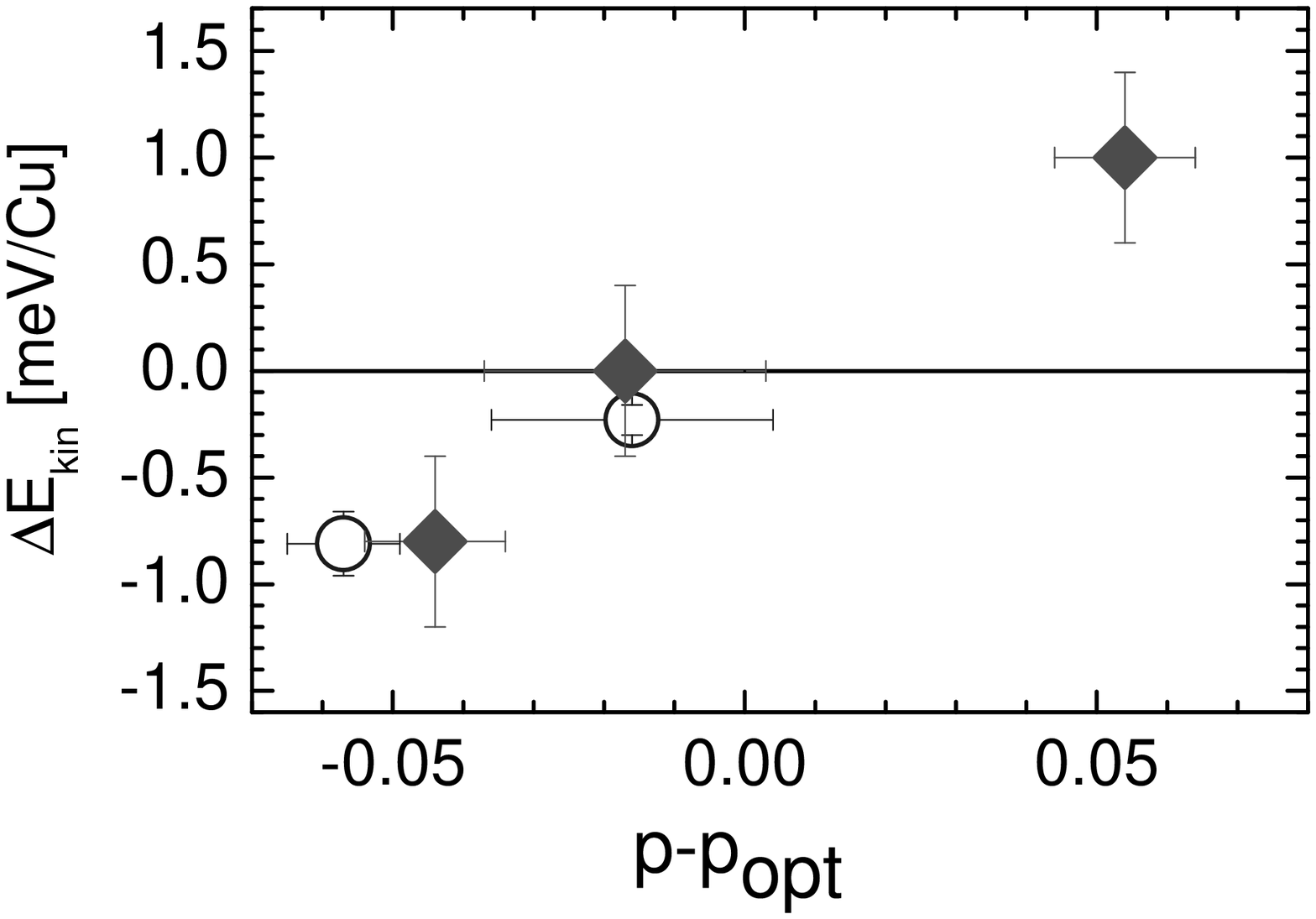}
  \caption{Change $\Delta E_{kin}$ of the KE, in meV per copper site
vs the charge $p$ per copper site with
respect to $p_{opt}$ [Eq.~\protect{\eqref{eq:31}}] in BSCCO.
Full diamonds: data from
Ref.~\protect{\onlinecite{ref47}}, high-frequency cutoff $1\,$eV. Open
circles: data from Ref.~\protect{\onlinecite{ref48}}, high frequency
cutoff $1.25\,$eV. Error bars: vertical, uncertainties due to the
extrapolation of the temperature dependence of the normal state spectral
weight down to zero temperature; horizontal, uncertainties resulting
from $T_c/T_{c,max}$ through Eq.~\protect{\eqref{eq:31}}.
Deutscher {\it et al.}\protect{\cite{ref87a}} took
$T_{c,max} = (83\pm2)\,$K for films and $(91\pm2)\,$K for crystals.
Adapted from Ref.~\protect{\onlinecite{ref87a}}.
}
  \label{fig:20}
\end{figure}
(not shown) the normal state temperature dependence is closer to
linear than quadratic and the underdoped sample shows a very
flat region before entering the superconducting state. This demonstrates
once more that there is as yet no strong consensus in the literature
as to the $T$ dependence of the normal state. (Please see also
Ref.~\onlinecite{ref89a}.)
The same paper also analyses the data in terms
of KE change for three samples UND70K, OPT80K, and OVR63K, as shown
in Fig.~\ref{fig:20} reproduced from Deutscher {\it et al.}\cite{ref87a}
The horizontal axis is $p-p_{opt}$ where $p$ is the charge per Cu atom related
to $T_c$ by
\begin{equation}
  \label{eq:31}
  \frac{T_c}{T_{c,opt}} = 1-86.2\left(p-p_{opt}\right)^2,
\end{equation}
with $T_{c,opt}$ the maximum critical temperature for the Bi2212 series.%
\cite{ref89} What is clear from this figure is that there is a smooth
crossover from standard behavior on the overdoped side to anomalous
behavior on the underdoped side. The very recent data of
Carbone {\it et al.}\cite{ref33a} lends further support to this
conclusion.

\subsection{Cluster dynamical mean field of the $t$--$J$ model}
\label{ssec:4d}

Recently Haule and Kotliar\cite{ref89b} calculated the optical
conductivity of the $t$--$J$ model within a cluster DMFT (CDMFT).
(Please see also earlier work by Maier {\it et al.}\cite{ref89c}
\begin{figure}[tp]
%  \centering
%  \includegraphics[width=12cm]{srFig32.eps}
  \includegraphics[width=8cm]{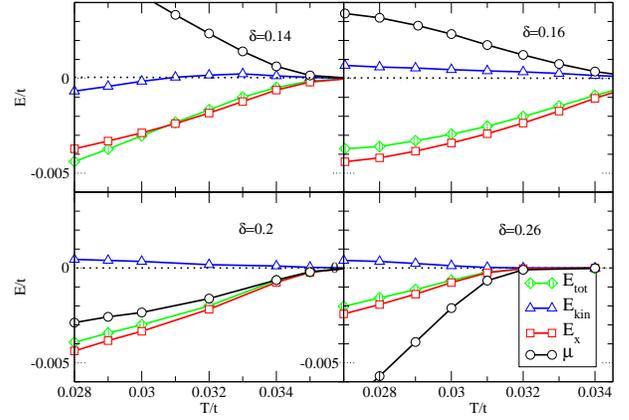}
  \caption{(Color online)
The difference between the superconducting and normal state
energies as a function of temperature. The following curves are shown:
up-triangles - $E_{kin-S}-E_{kin-N}$; squares - $E_{x-S}-E_{x-N}$;
diamonds - $E_{tot-S}-E_{tot-N}$; circles - $\mu_S-\mu_N$.
Adapted from Ref.~\protect{\onlinecite{ref89b}}.
}
  \label{fig:32}
\end{figure}
based on the Hubbard model.) Haule and Kotliar\cite{ref89b} treat
the temperature and doping
dependence and address the issue of the change in KE of the holes
when superconductivity sets in. They find, in agreement with
experiment, that on the overdoped side the KE makes a negative
contribution to the condensation energy as in conventional BCS theory
but that there is a crossover to the opposite case on the underdoped
side. This is shown in Fig.~\ref{fig:32} reproduced from
Ref.~\onlinecite{ref89b}. What is shown is the difference between the
superconducting and normal state energies $(E)$ as a function of
temperature, both in units of $t$ (nearest neighbor hopping). The
up-triangles give the KE contribution $(E_{kin})$,
the squares are the super exchange $(E_x)$,
and the diamonds the total energy $(E_{tot})$.
We see a change in sign in the KE
contribution upon condensation as we go from the overdoped to
the underdoped regime. ($\delta$ denotes the doping.) These results
contrast with earlier work of Maier {\it et al.}\cite{ref89c} also based
on the Hubbard model. There, the KE is found to be lower in the
superconducting state than in the normal state for both values
of doping presented, namely $\delta=0.05$ and $\delta=0.2$ (overdoped).
In the underdoped case the potential energy increases slightly
above its normal state value below $T=T_c$. For the overdoped case
it does decrease very slightly but plays a much smaller role in the
condensation than the corresponding drop in KE.

We end this section by mentioning two related works\cite{ref127,ref128}
based on the negative $U$ Hubbard model which has been used
successfully to describe the BCS - Bose Einstein (BE) crossover.
Toschi {\it et al.}\cite{ref127} employ DMFT and Kyung {\it et al.}%
\cite{ref128} a cellular DMFT and obtain very similar results. Both
normal and superconducting states are considered. Both groups find
a change of sign in the KE difference between superconducting and
\begin{figure}[tp]
  \centering
  \includegraphics[width=9cm]{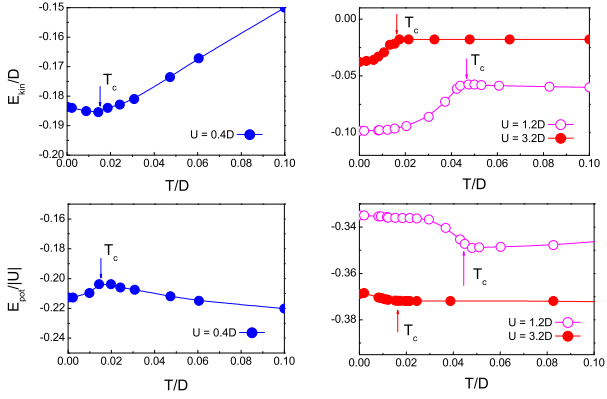}
  \caption{(Color online) Low-temperature behavior of the normalized kinetic
energy $E_{\rm kin}/D$ (upper panels) and the normalized potential energy
$E_{\rm pot}/\vert U\vert$ (lower panels) as a function of the normalized
temperature $T/D$. Here $U$ is the Hubbard attraction.
The critical temperature is marked by arrows. Adapted from
Ref.~\protect{\onlinecite{ref127}}.}
  \label{fig:32a}
\end{figure}
normal state as one goes from underdoping (anomalous) to overdoping
(conventional BCS behavior) as seen in Fig.~\ref{fig:20}. In
Fig.~\ref{fig:32a} we reproduce from Ref.~\onlinecite{ref127} results
for KE $(E_{\rm kin})$ in units of the band width $D$ vs the
normalized temperature $T/D$ (upper frames) and for the potential energy
$E_{\rm pot}$ in units of $\vert U\vert$ (lower frames). Three values
of $U$ are considered, $U=0.4\,D$ in the left frames and $U=1.2\,D$
and $3.2\,D$ in the frames on the right. These values correspond to
BCS, intermediate, and BE (Bose - Einstein) regimes, respectively. The frames
on the left show conventional behavior with a small increase in KE
in the superconducting state and a decrease in potential energy.
On the other hand, in the frames on the right the KE decreases
while the potential
energy increases which is referred to as KE driven superconductivity.

\section{Models of the Pseudogap State}
\label{sec:5}
\subsection{The Preformed Pair Model}
\label{ssec:5a}

Another model for superconductivity in the oxides is the preformed
pair model.\cite{ref91,ref92,ref93,ref94,ref95,ref96} The idea is
that the pairs form at a temperature $T^\ast$, the pseudogap formation
temperature. The superconducting state emerges from the preformed
pair state at a lower temperature $T_c$ when phase coherence sets in. The
pseudogap regime is then due to superconducting phase fluctuations.
In a recent paper Eckl {\it et al.}\cite{ref97}
considered the effect of phase fluctuations on the OS, i.e.: the KE
in the pseudogap regime. [See also the related work of
Kope\'c, Ref.~\onlinecite{ref97a}.]
They start with a Hamiltonian which contains
two terms, the KE
\begin{equation}
  \label{eq:35}
  K = -t\sum\limits_{\langle ij\rangle\sigma}\left(
  c^\dagger_{i\sigma}c_{j\sigma}+c^\dagger_{j\sigma}c_{i\sigma}\right)
\end{equation}
and a pairing term
\begin{equation}
  \label{eq:36}
  -\frac{1}{4}\sum\limits_{i\delta}\left(\Delta_{i\delta}\left\langle
  \Delta^\dagger_{i\delta}\right\rangle+\Delta^\dagger_{i\delta}
  \left\langle\Delta_{i\delta}\right\rangle\right),
\end{equation}
where $\delta$ connects the site $i$ to its nearest neighbors only and
$\langle ij\rangle$, as before, is limited to nearest neighbors as well.
Furthermore,
\begin{equation}
  \label{eq:37}
  \Delta^\dagger_{i\delta} = \frac{1}{\sqrt{2}}\left(c^\dagger_{i\uparrow}
  c^\dagger_{i+\delta\downarrow}-c^\dagger_{i\downarrow}
  c^\dagger_{i+\delta\uparrow}\right).
\end{equation}
Its average $\left\langle\Delta^\dagger_{i\delta}\right\rangle\equiv\Delta\,
\exp\left(i\Phi_{i\delta}\right)$ with $\Delta$ the gap amplitude is
assumed to have $d$-wave symmetry and the phase
\begin{equation}
\label{eq:38}
\Phi_{i\delta} = \begin{cases}
    \left(\phi_i+\phi_{i+\delta}\right)/2 & \mbox{for $\delta$ along $x$},\\
    \left(\phi_i+\phi_{i+\delta}\right)/2+\pi &
     \mbox{for $\delta$ along $y$}.
\end{cases}
\end{equation}
The phases are assumed to fluctuate according to classical XY free energy.
In this model the Kosterlitz-Thouless transition $T_{KT}$ is identified with
\begin{figure}[tp]
%  \centering
  \includegraphics[width=8cm]{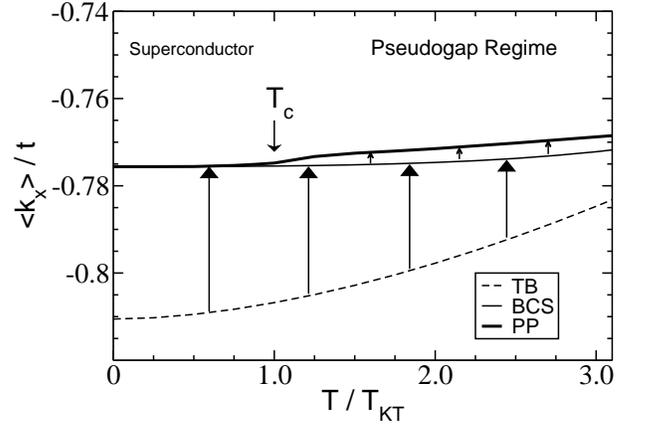}
  \caption{Kinetic energy per bond, $\langle k_x\rangle$, as a function of
temperature for non interacting tight-binding electrons (TB), the BCS
solution (BCS), and the phase fluctuation (PP) model for $\mu=0$
$(\langle n\rangle = 1)$. The large vertical arrows indicate the increase
in KE upon pairing, relative to the free tight-binding model, and
the small arrows indicate the additional increase due to phase
fluctuations. This additional phase fluctuation energy rapidly
vanishes near $T_c=T_{KT}$, which causes the significant change in the
OS upon entering the superconducting state at $T_{KT}=0.1\,t$. Note
that the thick line follows the actual KE encountered in this model, when
going from the pseudogap to the superconducting regime. Adapted from
Ref.~\protect{\onlinecite{ref97}}.
}
  \label{fig:22}
\end{figure}
the superconducting transition temperature $T_c$ and the mean field
temperature $T_{MF}$ at which the pairs form, is identified with the
pseudogap temperature $T^\ast$. They take $T_{KT}\simeq T_{MF}/4$ with
$T_{KT}=0.1\,t$. In Fig.~\ref{fig:22}, reproduced from Ref.~\onlinecite{%
ref97}, we show the KE per bond, $\langle k_x\rangle$ as a function
of the reduced temperature $T/T_{KT}$. Here
\begin{equation}
  \label{eq:39}
   \langle k_x\rangle = -t\sum\limits_\sigma\left\langle
   c^\dagger_{i\sigma}c_{i+x,\sigma}+c^\dagger_{i+x.\sigma}c_{i\sigma}
   \right\rangle.
\end{equation}
The dashed curve is the result of the simple tight-binding
Hamiltonian [$K$ of Eq.~\eqref{eq:35}
only] for zero chemical potential and $\langle n\rangle=1$. The light
solid line is the result of BCS mean field and the heavy solid
line the result obtained by taking into account phase fluctuations
within the preformed pair model. The mean field BCS condensation
increases the KE above its
tight-binding value and the phase fluctuations provide an additional
KE increase which vanishes at $T_c=T_{KT}$ and this causes significant
\begin{figure}[tp]
%  \centering
  \includegraphics[width=8cm]{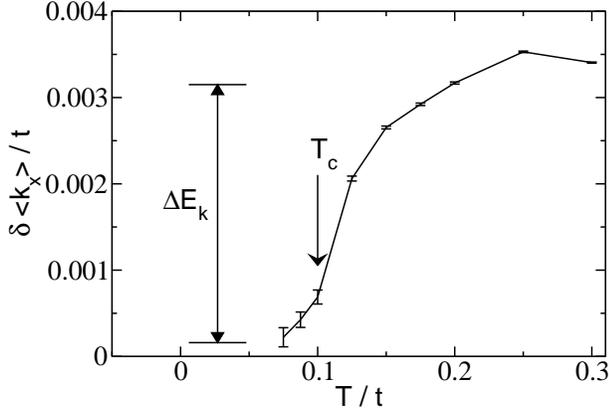}
  \caption{Kinetic energy contribution from phase fluctuations
$\delta\langle k_x\rangle \equiv \langle k_x\rangle_{PP}-
\langle k_x\rangle_{BCS}$. One can clearly see the sharp decrease of KE
near the Kosterlitz-Thouless transition $T_{KT}=0.1\,t\equiv T_c$.
$\Delta E_k$ gives an estimate of the kinetic condensation energy.
Adapted from Ref.~\protect{\onlinecite{ref97}}.
}
  \label{fig:23}
\end{figure}
change in the OS as the superconducting state is entered at
$T_{KT}=0.1\,t$ in this model calculations. The KE gain from the
phase fluctuations $\delta\langle k_x\rangle \equiv \langle k_x\rangle_{PP}-
\langle k_x\rangle_{BCS}$ is shown in Fig.~\ref{fig:23} which shows
a sharp decrease in KE near the Kosterlitz-Thouless transition and
$\Delta E_k$ gives an estimate of the KE condensation $\propto 0.003\,t$
of the same order as was found in other models. However, this is not the
KE change between normal and superconducting state at $T=0$
as discussed previously. 
Rather it is a change in KE due to the suppression of phase
fluctuations present in the pseudogap state. At $T_c=T_{KT}$ the
phases become locked in and $\delta\langle k_x\rangle$ consequently
becomes zero. The change in KE on formation of the Cooper pairs is
never explicitly considered in this model as the pairs are assumed to
form at a much higher temperature $T=T^\ast = 0.4\,t$.

\subsection{The $D$-density Waves, Competing Order Model}
\label{ssec:5b}

There are other very different models that have been proposed for the
pseudogap state which exists at temperatures above the superconducting
state in the underdoped regime. One proposal is $D$-density waves\cite{%
ref98,ref99,ref100,ref101,ref102,ref103,ref103a,ref104,ref105,ref106}
(DDW) which
falls into the general category of competing interactions. The view is
that a new phase, not superconducting, but having a gap with $d$-wave
symmetry,
forms at $T^\ast$ and the superconductivity which arises only at some
lower temperature $T_c$ is based on this new ground state.\cite{%
ref98,ref99,ref100,ref101,ref102,ref103,ref103a,ref104,ref105,ref106,%
ref107,ref108,ref109,ref110,ref111,ref112,ref113,ref114} The DDW state breaks
time reversal symmetry because it introduces bond currents\cite{ref115,%
ref116} with associated small magnetic moments and a gap forms at the
antiferromagnetic Brillouin zone boundary. The mean field DDW Hamiltonian is
\begin{equation}
  \label{eq:40}
  H = \sum\limits_{{\bf k},\sigma}\left[\left(\epsilon_{\bf k}-
\mu\right) c^\dagger_{{\bf k}\sigma}c_{{\bf k}\sigma}+i D_{\bf k}
c^\dagger_{{\bf k}\sigma}c_{{\bf k}+{\bf Q}\sigma}\right],
\end{equation}
where the sum on {\bf k} ranges over the entire Brillouin zone and
{\bf Q} is the commensurate wave vector $(\pi/a,\pi/a)$. Here
$D_{\bf k} = D_0\left[\cos(k_xa)-\cos(k_ya)\right]/2$ with $D_0$ the
gap amplitude. Within a mean field approximation
the probability of occupation of
the state $\vert{\bf k}\sigma\rangle$ is given by\cite{ref112}
\begin{eqnarray}
  n_{{\bf k}\sigma}(T) &=& \frac{1}{2E_{\bf k}}\left\{E_{\bf k}\left[
  f(T,\xi_{+{\bf k}})+f(T,\xi_{-{\bf k}})\right]\right.\nonumber\\
  &&\left.+\epsilon_{\bf k}\left[
 f(T,\xi_{+{\bf k}})-f(T,\xi_{-{\bf k}})\right]\right\},
  \label{eq:41}
\end{eqnarray}
where $\xi_{\pm{\bf k}} = -\mu\pm E_{\bf k}$ and $E_{\bf k} = \sqrt{%
\epsilon_{\bf k}^2+D_{\bf k}^2}$. The number density is
\begin{equation}
  \label{eq:42}
  n(T) = \frac{2}{N}\sum\limits_{{\bf k}\in{\rm MBZ}}\left[f(T,\xi_{-{\bf k}})+
  f(T,\xi_{+{\bf k}})\right],
\end{equation}
where the sum over {\bf k} is restricted to half the original
Brillouin zone, i.e.: to the magnetic Brillouin zone (MBZ).
The KE from Refs. \onlinecite{ref108}, \onlinecite{ref113}, and
\onlinecite{ref114} is denoted by $W(D,T)$ and is
\begin{equation}
  \label{eq:43}
  W(D,T) = \frac{2}{N}\sum\limits_{{\bf k}\in{\rm MBZ}}
  \frac{\epsilon_{\bf k}^2}
  {E_{\bf k}}\left[f(T,\xi_{+{\bf k}})-f(T,\xi_{-{\bf k}})\right].
\end{equation}
Here and later in this section we give formulas only for the simpler
case of $t'=0$, i.e.: only nearest neighbors. The results to be presented,
however, are based on straightforward generalizations that include
second nearest neighbors hopping.

\begin{figure}[tp]
\vspace*{-4mm}
  \includegraphics[width=9cm]{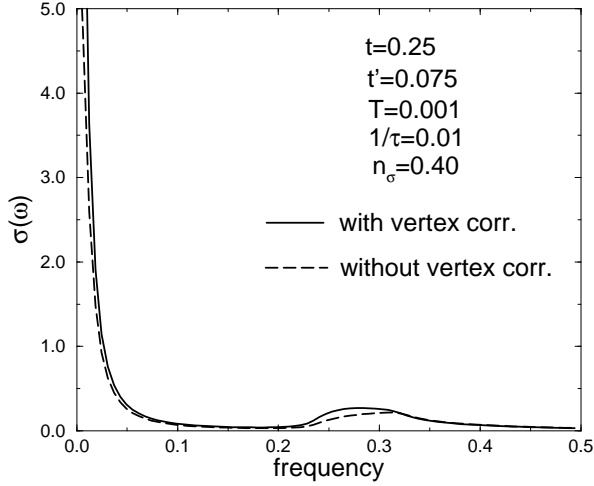}
  \caption{Optical conductivity $\sigma(\omega)$ with (solid line)
and without (dashed line) vertex corrections.
Adapted from Ref.~\protect{\onlinecite{ref119}}.
}
  \label{fig:25}
\end{figure}

Aristov and Zeyher\cite{ref119} calculated the optical conductivity in
the DDW model with and without vertex corrections to the usual current
operator. The previous work of Valenzuela {\it et al.}\cite{ref110} was
without vertex corrections as is the more recent work of Gerami and
Nayak\cite{ref120} who, however, consider the effect of anisotropic
scattering. These authors use a current operator which includes a term
proportional to the gap velocity $\nabla_{\bf k}D_{\bf k}$ which is
\begin{figure}[tp]
\vspace*{-4mm}
  \includegraphics[width=9cm]{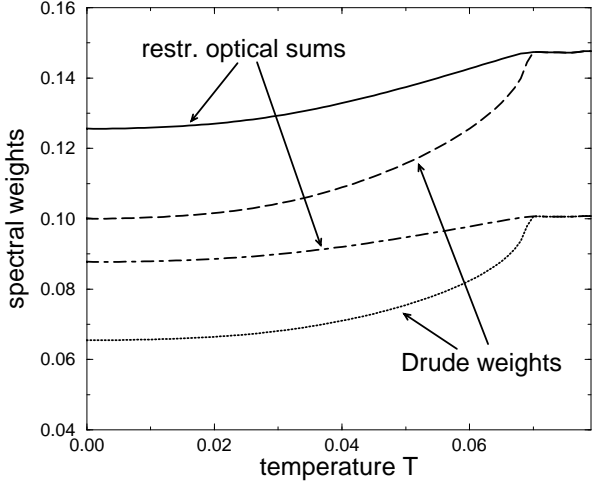}
  \caption{Restricted optical sums with (solid line) and without
(dot-dashed line) vertex corrections, and Drude weights with (dashed
line) and without (dotted line) vertex corrections.
Adapted from Ref.~\protect{\onlinecite{ref119}}.
}
  \label{fig:24}
\end{figure}
introduced so as to ensure charge conservation as is discussed in more
detail by Benfatto {\it et al.}\cite{ref113} Here we follow
Aristov and Zeyher\cite{ref119} and show in Fig.~\ref{fig:24} their
results for the optical spectral weight as a function of temperature
$T$ in units of the hopping parameter $t$. The solid curve is with
and the dash-dotted curve is without vertex corrections. 
We see that while vertex corrections
have changed the magnitude of the OS they have changed much less its
temperature dependence which is somewhat more pronounced in the solid
curve. In both cases the OS decreases with decreasing temperature as
was the case in BCS theory. Also shown in Fig.~\ref{fig:24} are results
for the Drude weight separately, dashed and dotted lines with and
without vertex corrections. We see that it is also strongly enhanced by
vertex corrections. In Fig.~\ref{fig:25} we reproduce
results of Aristov and Zeyher\cite{ref119} for the conductivity
$\sigma(\omega)$ vs frequency $\omega$ for $t=0.25\,$eV, $t'=0.076$,
$T=0.001$, $n=0.4$ and a scattering rate of $1/\tau = 0.01$. It is
seen that both the Drude region (intra-band) and the inter-band
transitions which set in at higher $\omega\ge 0.2$ are larger when
vertex corrections are included but that beyond $\omega\ge0.3$ the two
curves merge and the vertex corrections are no longer important.

In the continuum limit of the DDW model Valenzuela {\it et al.}%
\cite{ref110} have obtained useful analytic results for intra
(Drude) and inter (verticle transitions) separately. Vertex
corrections are neglected and the limit of zero impurity scattering
was taken. They find
\begin{equation}
  \label{eq:46}
  \sigma_{intra}(T,\omega) = -\delta(\omega)\pi e^2
  \int\limits_{-\infty}^\infty\!d\nu\,\left[\frac{\partial f(t,\nu)}
  {\partial \nu}\right]g_1(\nu),
\end{equation}
with
\begin{equation}
  \label{eq:47}
  g_1(\nu) = (\hbar v_F)^2 N(0)\frac{2}{\pi}\int\limits_0^{\pi/2}\!
  d\theta\,\Re{\rm e}\sqrt{1-\frac{D_0^2}{(\mu+\nu)^2}\cos^2\theta},
\end{equation}
which can be written in terms of elliptic integrals. Here, $v_F$
is the Fermi velocity. Note
also that the Fermi factor $f(T,\nu)$ has $\mu=0$ and the chemical
potential has been transfered to $g_1(\nu)$. For the inter-band
\begin{widetext}
\begin{equation}
  \label{eq:48}
  \sigma_{inter}(T,\omega) = (\hbar v_F)^2N(0)\frac{1}{\pi}\left(\frac{2D_0}
  {\omega}\right)^2\int_0^{\pi/2}\!d\theta\,\Re{\rm e}
  \left[\frac{\cos^2\theta}{\sqrt{1-\left(\frac{2\Delta}{\omega}\right)^2
  \cos^2\theta}}\right]B(T,\omega),
\end{equation}
\end{widetext}
where $B(T,\omega)$ is a universal thermal factor
\begin{equation}
  \label{eq:49}
  B(T,\omega) = \frac{\pi e^2}{\omega}\frac{\sinh(\beta\omega/2)}
  {\cosh(\beta\mu)+\cosh(\beta\omega/2)},\qquad \beta=\frac{1}{k_B T},
\end{equation}
which at $T=0$ becomes proportional to a theta function
$\theta(\omega-2\vert\mu\vert)$. This provides a low energy cutoff to the
inter-band transitions. When impurity scattering is included the
delta function $\delta(\omega)$ in Eq.~\eqref{eq:46} broadens into a
Lorentzian and temperature leads to the overlap of the two contributions
as seen in Fig.~\ref{fig:25}.

\begin{figure}[tp]
%  \centering
%  \includegraphics[width=10cm,angle=270]{srFig26.eps}
  \includegraphics[width=6cm,angle=270]{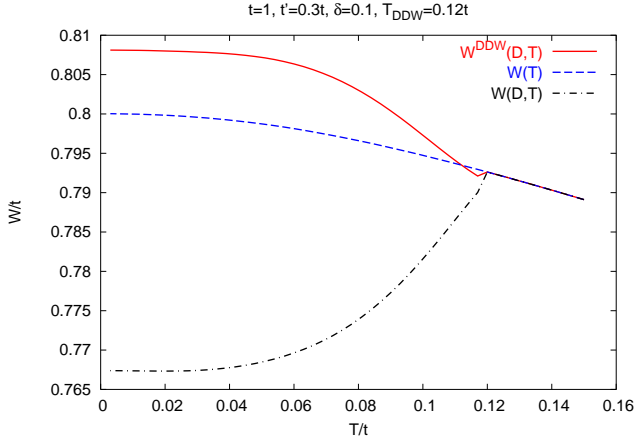}
  \caption{(Color online)
Optical spectral weight in the presence of a $t'$ term in the band
dispersion.
Here $W(D,T)$ (Eq.~\protect{\eqref{eq:43}}, dash-dotted line),
$W^{DDW}(D,T)$ (Eq.~\protect{\eqref{eq:49a}} (solid line) are shown
together with $W(T)$ (dashed line) which is for the non interacting
case. The parameters are $t'=0.3\,t$, $T_{DDW}=0.12\,t$,
$D(0) = 4T_{DDW}$, and the doping $\delta = 0.1$. The chemical potential is
evaluated self consistently at each temperature.
Observe that near $T_{DDW}$ a small decrease of $W^{DDW}$ with
respect to $W(T)$ is seen, due to the change of chemical potential
near $T_{DDW}$. Adapted from Ref.~\protect{\onlinecite{ref113}}.
}
  \label{fig:26}
\end{figure}
We note that the kinetic energy shows the same trend in its temperature
dependence as does the OS of Fig.~\ref{fig:24}. This is shown in
Fig.~\ref{fig:26} which we took from the work of Benfatto
{\it et al.}\cite{ref113} and where it is denoted by $W(D,T)$ as
the dash-dotted curve for the parameters shown in the figure and described
in the caption. We see the same decreasing trend with decreasing
temperatures as for the OS. There are two other curves, the dashed
one is for
a pure tight-binding band with no DDW and is included for comparison.
The solid curve is for $W^{DDW}(D,T)$ given by
\begin{equation}
  \label{eq:49a}
  W^{DDW}(D,T) = -\frac{1}{N}\sum\limits_{\bf k}^{MBZ}E_{\bf k}
  \left[f(T,\xi_{+{\bf k}})-f(T,\xi_{-{\bf k}})\right]
\end{equation}
(again in units of $\pi e^2/\hbar^2$)
in the notation of Benfatto
{\it et al.}\cite{ref113,ref114} and requires explanation. It was obtained
from a current operator which was modified to ensure charge
conservation without the need for vertex corrections. This current
operator has been used in other works\cite{ref108,ref120} as well.
As noted by Aristov and Zeyher\cite{ref119}, however, such a
procedure tends to overestimate the conductivity at higher
frequencies and the OS now shows an increase as $T$ is decreased.

We can also add a mean field BCS term
\begin{equation}
  \label{eq:44}
  H' = \sum\limits_{\bf k}\left[\Delta^\ast_{\bf k}c_{-{\bf k}\downarrow}
  c_{{\bf k}\uparrow}+ h.c.\right],
\end{equation}
to the Hamiltonian \eqref{eq:40},
where $\Delta_{\bf k}$ is the superconducting order parameter.
$\Delta_{\bf k}=\Delta_0\left[\cos(k_xa)-\cos(k_ya)\right]$ with
$\Delta_0$ the superconducting gap amplitude. In this case the
OS in units of $\pi e^2/\hbar^2$ is given by
\begin{eqnarray}
  W^{DDW}(D,\Delta,T) &=& \frac{2}{N}
  \sum\limits_{\bf k}^{\rm MBZ} E_{\bf k}\left[
  \frac{\xi_{+{\bf k}}}{E_{+{\bf k}}}\tanh\left(\frac{E_{+{\bf k}}}{2T}\right)
  \right.\nonumber\\
  &&\left.
 -\frac{\xi_{-{\bf k}}}{E_{-{\bf k}}}\tanh\left(\frac{E_{-{\bf k}}}{2T}\right)
  \right],
  \label{eq:45}
\end{eqnarray}
with $E_{\pm{\bf k}} = \sqrt{\xi^2_{\pm{\bf k}}+\Delta^2_{\bf k}}$.

\begin{figure}[tp]
%  \centering
%  \includegraphics[width=14cm]{srFig27.eps}
  \includegraphics[width=8cm]{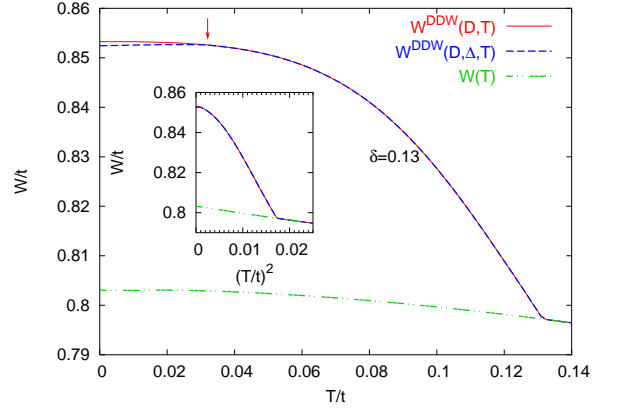}
  \caption{(Color online)
Optical spectral weight in units of $e^2\pi/\hbar^2$ in the
normal state
(dash double-dotted line), in the DDW state
Eq.~\protect{\eqref{eq:49a}} (solid line)
and in DDW+SC state Eq.~\protect{\eqref{eq:45}} (dashed line).
The values of parameters are
for the doping $\delta = 0.13$ [$D_0(T=0)=0.92\,t$,
$\Delta(T=0)=0.064\,t$, see Refs.~\protect{\onlinecite{ref107}} and
\protect{\onlinecite{ref108}}
for further details. The critical temperature is marked by the arrow.
Observe that the decrease of $W^{DDW}(D,\Delta,T)$ below $T_c$ is
small. Inset: spectral weight plotted as a function of
$(T/t)^2$. Adapted from Ref.~\protect{\onlinecite{ref114}} (see also
Ref.~\onlinecite{ref108}).
}
  \label{fig:27}
\end{figure}
In Fig.~\ref{fig:27} we show results reproduced from Benfatto and
Sharapov\cite{ref114} for the effect of combined DDW and superconducting
transition described by Eq.~\eqref{eq:45}.
The optical weight is given in units of $t$ (nearest neighbor hopping)
as is the temperature. The dash-double-dotted curve $[W(T)]$ is for
reference
and gives results for the tight binding band without interactions while
the solid lines gives $W^{DDW}(D,T)$ as before and the dashed
curve is for $W^{DDW}(D,\Delta,T)$ which includes the effect of a $d$-wave
superconducting gap of amplitude $\Delta$. We note that, as expected,
this leads to an increase in KE with respect to the pure DDW case and,
therefore, a drop in the OS.

It is clear that, as yet, a complete theory of the OS in the DDW model
does not
exist. The calculations of Aristov and Zeyher\cite{ref119} with vertex
corrections properly accounted for give the opposite temperature
dependence than found experimentally. This theory, however, does not
include correlations beyond those directly responsible for the DDW
transition. It is clear from what we have described here that correlations
leading to lifetime effects need to be included as these are very
closely related to the observed temperature dependence of the OS
and cannot be ignored.

\section{Optical Spectral Weight Distribution}
\label{sec:6}

In this review we focused mainly on the OS for a single band
\begin{figure}[tp]
%  \centering
  \includegraphics[width=8cm]{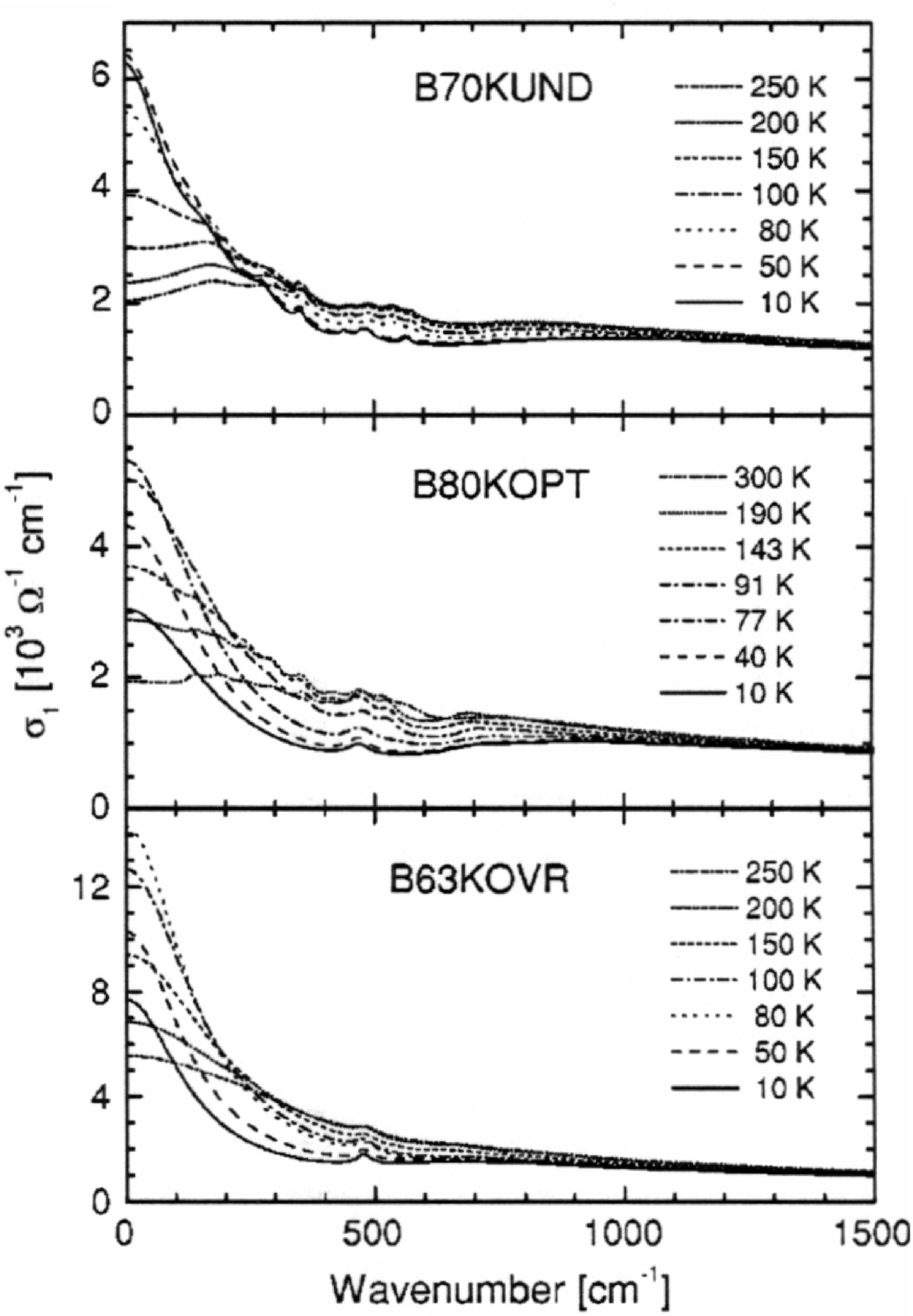}
  \caption{Selection of conductivity spectra for
three BSCCO samples,  underdoped, $T_c=70\,$K (B70KUND, top frame),
optimally doped, $T_c=80\,$K (B80KOPT, middle frame), and overdoped,
$T_c=63\,$K (B63KOVR, bottom frame). For the
B70KUND sample, within the showed spectral range, the spectra at
$50\,$K and $10\,$K are indistinguishable.
 Adapted from Ref.~\protect{\onlinecite{ref47}}.
}
  \label{fig:28}
\end{figure}
integrated over all energies of relevance. As is seen from Eq.~\eqref{eq:1}
this quantity can be computed from a knowledge of the single electron
spectral density $A({\bf k},\omega)$ which determines the probability
of occupation $n_{{\bf k},\sigma}$ of the state $\vert{\bf k},\sigma\rangle$.
This is a much simpler problem than computing the frequency dependent
conductivity from a Kubo formula which involves the two-particle Green's
functions. However, calculating the full frequency dependent conductivity
cannot be avoided if one wishes to discuss the optical spectral weight
distribution. The partial integration of
$\sigma_1(T,\omega)$ to a maximum $\omega$ equal to $\omega_M$ has
proved be very useful and has provided valuable information
about normal and superconducting state beyond what is obtained from the
OS itself. We give here only two
examples. In Fig.~\ref{fig:28} we reproduce the conductivity
data of Santander-Syro
{\it et al.}\cite{ref47} in three BSCCO samples, underdoped $T_c=70\,$K
(B70KUND), optimally doped at $T_c=80\,$K (B80KOPT), and overdoped
at $T_c=63\,$K (B63KOVR) at the various temperatures noted in the figure.
From these data it is possible to calculate
$W(T,\omega_M)=\int_{0^+}^{\omega_M}%
d\omega\,\sigma_1(T,\omega)$ at various temperatures.
Here the $0^+$ indicates that
\begin{figure}[tp]
%  \centering
  \includegraphics[width=8cm]{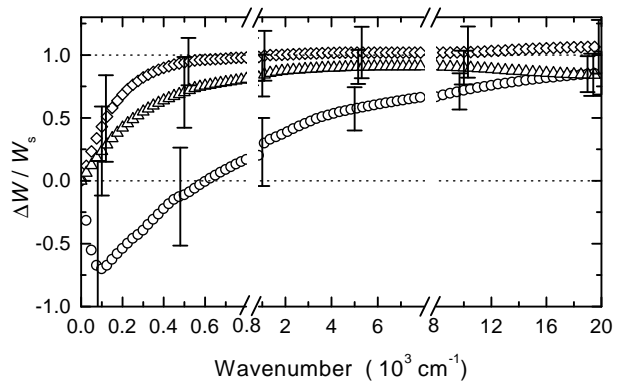}
  \caption{Rato $\Delta W/W_S$ vs frequency showing the exhaustion of the
Ferrell-Glover-Tinkham sum rule at conventional energies for the OVR (diamonds)
and OPT (triangles) samples. An
unconventional ($\sim 16 000\,$cm$^{-1}$ or $2\,$eV) energy scale is
required for the UND sample (circles). Note that the frequency scale changes at
800 and $8 000\,$cm$^{-1}$. The changes in spectral weight are taken
between $80\,$K - $10\,$K, $91\,$K - $10\,$K, and $100\,$K - $10\,$K
for the OVR, OPT, and UND samples, respectively. Adapted from
Ref.~\protect{\onlinecite{ref47}}.
}
  \label{fig:29}
\end{figure}
the delta function representing the  condensate has been left out.
In Fig.~\ref{fig:29}
we reproduce the experimental results of Santander-Syro {\it et
  al.}\cite{ref47} for the normalized change in spectral weight
$\Delta W/W_S$ vs $\omega_M$ in $10^3\,$cm$^{-1}$ based on the data of
Fig.~\ref{fig:28}. Open diamonds,
triangles, and circles are for B63KOVR, B80KOPT, and B70KUND, respectively.
Note the two breaks in the horizontal scale for $\omega_M$ at 800 and
$8 000\,$cm$^{-1}$. The change in optical spectral weight is slightly
different
\begin{figure}[tp]
%  \centering
%  \includegraphics[width=12cm]{srFig30.eps}
  \includegraphics[width=9cm]{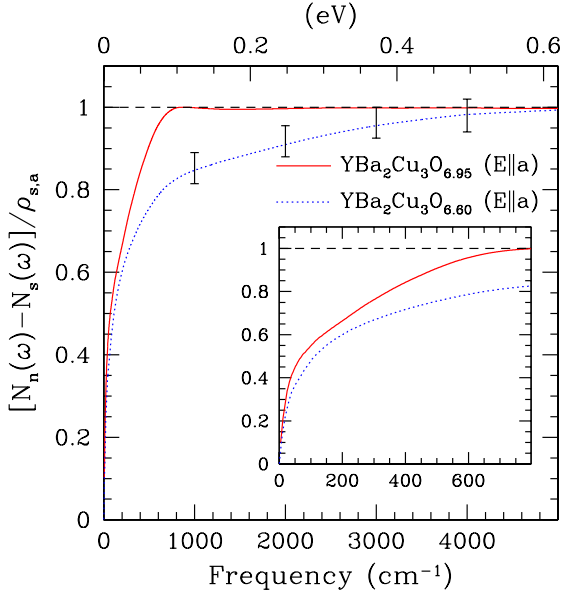}
  \caption{(Color online)
The normalized weight of the condensate [$N_n(\omega)-
N_s(\omega)]/\rho_{s,a}$ vs $\omega$ for optimally doped YBCO$_{6.95}$
(solid line)
and underdoped YBCO$_{6.60}$ (dotted line) along the $a$-axis direction.
The condensate for the optimally
doped material has saturated by $\simeq 800\,$cm$^{-1}$, while in the
underdoped material the condensate is roughly 80\% formed by this
frequency, but the other 20\% is not recovered until much higher
frequencies. The error bars on the curve for the underdoped material
indicate the uncertainty associated with the Ferrell-Glover-Tinkham
sum rule. Inset:
The low-frequency region. Adapted from Ref.~\protect{\onlinecite{ref121}}.
}
  \label{fig:30}
\end{figure}
for the OVR, OPT, and UND samples as indicated in the caption. Note
the approach to its asymptotic  value of one. For the overdoped case
the scale is $\sim 600\,$cm$^{-1}$ while for the UND sample it is
much larger and of order $2\,$eV. These experiments clearly reveal a
fundamental difference in behavior between overdoped and underdoped
samples. This was also observed in the YBCO series. Data from Homes
{\it et al.}\cite{ref121} are reproduced in Fig.~\ref{fig:30} for electric
field {\bf E} parallel to the $a$-axis in YBCO$_{6.95}$ (solid line)
and YBCO$_{6.60}$ (dotted line). For the optimally doped sample the OS
is rapidly saturated  ($\omega_M\simeq 800\,$cm$^{-1}$) while for the
underdoped case a frequency of about $9 000\,$cm$^{-1}$ is needed.

For a BCS $s$-wave superconductor the expectation is that the saturation
of the OS should occur at a frequency $\omega$ a few times the gap $\Delta$
even if
the system is dirty with scattering rates $\stackrel{>}{\sim}2\Delta$,
Refs.~\onlinecite{ref121} and \onlinecite{ref122}.
However, superconductors are better
described by Eliashberg theory which properly accounts for coupling
of the electrons to phonons. In this case the weight in the coherent
quasiparticle part of the spectral function is
$Z = 1/(1+\lambda)$ where $\lambda$ is the mass enhancement factor. The rest
of the spectral weight lies in an incoherent phonon induced band at
higher energy, usually in the infrared. This part of the spectral
function $A({\bf k},\omega)$ contributes the so called Holstein band to
the optical conductivity. Only the quasiparticle part is included in
BCS theory, yet for $\lambda > 1$, say, more of the electron spectral
weight is in the incoherent part than one finds in the coherent part.
This part introduces a new energy scale into the problem, namely, an
average phonon energy and it is no longer true to say that readjustment
of optical spectral weight on entering the superconducting state
can only occur on the scale of twice the gap. In fact, one should expect
that when the electrons pair, an absorption process involving a phonon
$\omega_E$ would be changed and shifted to an energy of $2\Delta+\omega_E$,
thus shifting the Holstein band to higher energies and, hence, the incoherent
part of the optical spectral weight in the superconducting state
extends to higher energies. This implies that
\begin{figure}[tp]
\vspace*{-4mm}
  \includegraphics[width=9cm]{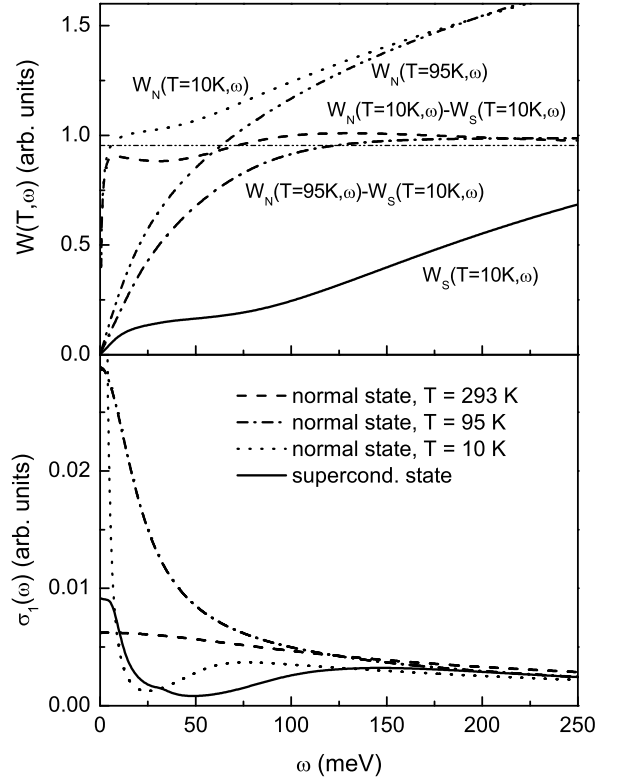}
  \caption{Top frame: Optical spectral weight $W(\omega,T) = \int_{0^+}^\omega\!
d\nu\,\sigma_1(\nu)$ for various cases as a function of $\omega$. The
dotted (dash-double-dotted) curve is for the normal state at $T=10\,$K
($T=95\,$K), the solid curve for the superconducting state at $T=10\,$K.
The dashed (dash-dotted) curve is the difference curve between superconducting
at $T=10\,$K and normal states
at $T=10\,$K ($T=95\,$K). The approach of the difference curve to its
saturated large $\omega$ value depends significantly on the temperature
used for the subtracted normal state. The thin dash-double-dotted
horizontal line is the value of the condensate contribution
(penetration depth). The bottom frame
shows the real part of the conductivity for the normal state at
$T=293\,$K (dashed curve), $T=95\,$K (dash-dotted curve), $T=10\,$K
(dotted curve), and for the superconducting state at a $T=10\,$K
(solid curve). All curves are for YBCO$_{6.95}$ with the impurity parameter
given in Ref.~\protect{\onlinecite{ref122a}}.
}
  \label{fig:31}
\end{figure}
$\int_{0^+}^\omega\!d\nu\,[\sigma_N(T,\nu)-\sigma_S(T,\nu)]$
for large $\omega$ should saturate from above rather than from below as
is the case in BCS. This is known from the $s$-wave phonon mediated
case\cite{ref129}
and is illustrated in Fig.~\ref{fig:31} taken from
Carbotte and Schachinger\cite{ref122a} for a $d$-wave superconductor
using for $I^2\chi(\omega)$ an MMP form. What is shown in the top frame are
numerical results for $W(T,\omega)$ defined as $W(T,\omega) =
\int_{0^+}^{\omega}\!d\nu\,\sigma_1(T,\nu)$. % where the $0^+$ indicates that
%the delta function representing the  condensate has been left out.
The real part of the conductivity $\sigma_1(T,\omega)$ in arbitrary
units on which these
various curves are based are shown in the bottom frame. These calculations
are all done for infinite bands and corresponding Eliashberg equations
for a $d$-wave superconductor so that the Ferrell-Glover-Tinkham
sum rule holds, i.e.: the total optical spectral weight is conserved
between normal and superconducting state.\cite{ref123,ref124} Parameters
were varied to get a good fit to data on YBCO$_{6.95}$. The reader is
referred to the paper of Schachinger and Carbotte\cite{ref33} for
details. Some impurity scattering in the unitary limit is included to
get Fig.~\ref{fig:31}. $W(T,\omega)$ is shown for
$\omega$ up to $250\,$meV. The solid curve is in the superconducting state at
$T=10\,$K and the dotted the normal state at the same temperature.
$W_N(T)$ (normal state) rises much more rapidly at small $\omega$
than does $W_S(T)$ (supercond. state) and goes
to much larger values. The difference $W_N(T=10\,{\rm K},\omega)-
W_S(T=10\,{\rm K},\omega)$ (dashed curve) is the amount of optical
spectral weight that has been transferred to the condensate between
$(0^+,\omega)$. This curve rapidly grows to a value slightly below
the horizontal line representing the condensate contribution to the
total sum rule. After this the remaining variation is small with a shallow
minimum around $30\,$meV followed by a broad peak around $60\,$meV which
falls above the thin dash-double-dotted horizontal line before gradually
falling again towards its asymptotic value which must be equal to the
condensate contribution. All these features can be understood from a
consideration of the curves for $\sigma_1(T,\omega)$ given in the
bottom frame. Comparing dotted and solid curves, we see that they cross
at three places on the frequency axis at $\omega_1\approx 8\,$meV,
$\omega_2\approx 32\,$meV, and $\omega_3\approx 130\,$meV. These
features are a result of the shift in incoherent background towards
higher energies due to the opening of the superconducting gap.

In an actual experiment it is usually not possible to access the
normal state at $T=10\,$K (say) so that $W_N(T=95\,{\rm K},\omega)$
just above $T_c = 92\,$K needs to be used (dash-double dotted curve).
The difference
$W_N(T=95\,{\rm K},\omega)-W_S(T=10\,{\rm K},\omega)$ is shown as the
dash-dotted curve in the top frame and is seen to merge with the
dashed curve only at higher values of $\omega$. Reference to the
bottom frame shows that the real part of the normal state conductivity
at $T=95\,$K (dash-dotted curve) is much broader than at
$T=10\,$K (dotted curve) and this accounts for the slower rise towards
saturation of the dash-dotted as compared to the dashed curve in the
top frame. We note, however, that
$W_N(T=95\,{\rm K},\omega)-W_S(T=10\,{\rm K},\omega)$ still approaches
the penetration depth curve from above but much of the structure seen
in the dashed curve is lost by using the data for the normal state at
$T=95\,$K rather than at $T=10\,$K. Nevertheless, the energy scale
over which the condensate is formed is set by the energy of the
spin fluctuation spectra which extends up to $400\,$meV in our
calculations and not by the gap value $2\Delta_0$.

%We give one more example of the usefulness of considering the spectral
%weight up to some finite, small $\omega_N$ rather than considering the
%total spectral weight for the particular band of interest. As we have
%said before, many experiments show that a pseudogap develops in
%underdoped systems. One might expect that its opening should lead to
%loose spectral weight at small frequencies as it opens.

\section{Summary}
\label{sec:7}

The temperature dependence of the OS has recently been measured up to
$\sim 300\,$K in the normal state of several cuprates as well as in
the superconducting state. Some controversy remains about details such
as the exact temperature dependence followed in the normal state.
However, it seems established that underdoped samples show anomalous
behavior when they enter the superconducting state. For a normal
BCS superconductor the KE should increase while it is seen to decrease
over its extrapolated normal state value. This, on its own, does not
mean that the condensation is purely driven by the kinetic energy but it
does mean that we are faced with non BCS behavior. For optimally
doped samples the change in KE is close to zero with a definite
crossover to conventional behavior in overdoped samples.
For the normal state many experiments, but not all, give a $T^2$ dependence
of the OS. Recent
DMFT calculations for the Hubbard model provide evidence for a
universal $T^2$ behavior in the normal state.
We argue, however, that such a law is not robust when
electron-boson theories for the interaction are considered. In this
case, if the boson energy is low, a linear in $T$ dependence can result and
by implication other dependences on $T$ can arise for different
electron-boson spectral densities. The same holds for coupling to
spin fluctuations in the NAFFL model. The issue of the temperature
dependence of the underlying normal state OS or KE impacts the
analysis of the change in KE that results from the superconducting
condensation. If the normal state KE decreases faster
at low temperatures than it does above $T_c$ (as we have seen in some
models) this could be interpreted as an anomalous superconducting
state, yet it would not be. On the other hand, in the
NAFFL model as well as in electronic models more generally, anomalous
behavior of the optical sum can be understood as due to the so called
collapse of the inelastic scattering rates in the superconducting
state if this reduces the effect of spin fluctuations at small
$\omega$. % In this model
%anomalous behavior of the OS can be understood on the basis of the
%so called collapse of the inelastic scattering rates as
%superconductivity sets in and reduces the effect of spin fluctuations
%at small $\omega$. Coupling to an optical resonance is also observed
%in some cases. 
Such a hardening of the spin fluctuation spectrum is
thought to be a viable explanation for the peak in the
microwave conductivity which is observed at a temperature
considerably below $T_c$. This effect, which can be modeled by a
low energy cutoff in the electron-spin fluctuation spectral density,
results directly in a decrease in KE of the underlying normal state
as additional coherence sets in with the establishment of
superconductivity. This idea is related to the work of
Norman and P\'epin\cite{ref78,ref79} who model directly the
charge carrier self
energy rather than go through the electron-boson spectral function.
They take the imaginary part of the self
energy $(\Sigma_2)$ to be large in the normal state (incoherence)
and increase coherence in the superconducting state through
a low frequency cutoff
below which $\Sigma_2$ is zero. Another related model, which is even simpler,
is to use a temperature dependent elastic scattering time which collapses
in the superconducting state as $\alpha T^4$ with $\alpha$ taken from
microwave experiments. All these mechanisms can easily provide savings in KE
sufficiently large to explain the OS experiments.
None, however, are fully quantitative at present. They also need to be
extended to the overdoped side of the phase diagram. In this case one
would expect the corresponding microwave peak to be greatly reduced
and be more conventional. More sophisticated,
but perhaps less transparent numerical calculations based on the
Hubbard model and on the $t$-$J$ model have also found that, as a
function of doping, the KE can favor pairing in underdoped systems
while for the overdoped case it behaves as for a conventional
superconductor. Earlier calculations based on the Hubbard model, however,
give kinetic energy pairing even on the overdoped side of the phase
diagram. By contrast, calculations based on the negative $U$ Hubbard model
used to describe the BCS-BE crossover do capture the observed change
in sign of the kinetic energy difference.

In Ref.~\onlinecite{ref122} and, more recently, in Ref.~\onlinecite{ref53}
the possibility is raised
that some of the effects studied in this review may arise from infinite
bands when a finite cutoff is applied to the optical sum integral as
it must in the analysis of experimental data. Certainly, for the high
$T_c$ oxides there is no clear `ending' of one band at a given energy
followed by a gap before the next band sets in at higher energies.
Instead, only a minimum is observed in the real part of the conductivity
as a function of $\omega$ at about $1.2\,$eV. This energy has been taken
as the cutoff on the transitions associated with the single band of
interest. This cutoff is also roughly consistent with what is known
about the width in energy of the bands in the oxides. Nevertheless,
ambiguity remains and a theoretical study of overlapping bands and what
this might imply for the single band sum rule would be valuable in
clarifying this situation further.

Much remains to be done to connect weak coupling
approaches with the more numerical strong coupling results. There
is also a need to get more accurate data which could resolve the
experimental debate that still goes on and to achieve quantitative
agreement with experiments.  What is clear, however, is that such
experiments can give useful information on the correlation effects
that are involved and they certainly have shown that
superconductivity on the underdoped side of the phase diagram of the
cuprates is likely to be unconventional. It cannot be understood on the basis
of a simple BCS model or an Eliashberg model with conventional phonons
as these are not expected to change much as superconductivity sets in
and they are distributed over a large energy range rather than peaked at low
$\omega$. However, we stress again that as we have seen in one case,
a small upturn in the
OS at low temperature in the superconducting state can arise in spin
fluctuation models without a readjustment of the electron-spin fluctuation
spectrum, when the spin fluctuation energy in the MMP spin
susceptibility is sufficiently small. While this upturn is less in the
superconducting state than in its normal state at this same temperature
(below $T_c$) it is above the value  extrapolated to $T=0$ from the
normal state above $T_c$. A second important conclusion of this review
is that lifetime effects are
important in determining the temperature dependence of the OS in
both normal and superconducting state. These have to be accounted for in
any definitive theory of this effect.

There is a need to continue to explore other models such as
phase fluctuations. In particular, the $D$-density wave model as a
possible competing order for the pseudogap phase has not yet included
effects of correlations beyond those that lead to the DDW order,
i.e: lifetime effects which we have argued to be central to this
problem. More work involving other models of interaction effects
so as to understand what other temperature laws are possible for the
OS vs $T$ would also be useful.\\[3ex]

\begin{acknowledgments}
Research supported in part by the Natural Sciences and Engineering
Research Council (NSERC) of Canada and the Canadian Institute of Advanced
Research (CIAR). We thank, E. Arrigoni, L. Benfatto, N. Bontemps,
R.P.S.M. Lobo, D. van der Marel, F. Marsiglio,
E. Nicol, S. Sharapov, A. Toschi, and R. Zeyher for discussion
and/or correspondence.
\end{acknowledgments}

\clearpage
%\begin{figure}[tp]
%  \centering
%  \includegraphics[width=10cm]{srFig31.eps}
%  \caption{Top frame: Optical spectral weight $W(\omega,T) = \int_{0^+}^\omega\!
%d\nu\,\sigma_1(\nu)$ for various cases as a function of $\omega$. The
%dotted (dash-double-dotted) curve is for the normal state at $T=10\,$K
%($T=95\,$K), the solid curve for the superconducting state at $T=10\,$K.
%The dashed (dash-dotted) curve is the difference curve between superconducting
%at $T=10\,$K and normal states
%at $T=10\,$K ($T=95\,$K). The approach of the difference curve to its
%saturated large $\omega$ value depends significantly on the temperature
%used for the subtracted normal state. The thin dash-double-dotted
%horizontal line is the value of the condensate contribution
%(penetration depth). The bottom frame
%shows the real part of the conductivity for the normal state at
%$T=293\,$K (dashed curve), $T=95\,$K (dash-dotted curve), $T=10\,$K
%(dotted curve), and for the superconducting state at a $T=10\,$K
%(solid curve). All curves are for YBCO$_{6.95}$ with the impurity parameter
%given in Ref.~\protect{\onlinecite{ref122a}}.
%}
%  \label{fig:31}
%\end{figure}

\begin{thebibliography}{999}
\bibitem{ref88}This was pointed out in footnote 15 of
  Ref.~\onlinecite{ref87a}.
\end{thebibliography}
\end{document}